\documentclass[12pt,preprint]{emulateapj}
\usepackage{amssymb,amsmath}
\usepackage{color,hyperref}
\definecolor{linkcolor}{rgb}{0,0,0.25}
\hypersetup{
  colorlinks=true,        
  linkcolor=linkcolor,    
  citecolor=linkcolor,    
  filecolor=linkcolor,    
  urlcolor=linkcolor      
}
\newcounter{address}
\newcommand{\ie}{i.e.}
\newcommand{\etal}{et al.}
\newcommand{\dd}{\mathrm{d}}
\newcommand{\eg}{e.g.}
\newcommand{\vs}{vs.}
\newcommand{\eqnname}{Equation}

\newcommand{\sectionname}{$\mathsection$}
\newcommand{\inv}{\ensuremath{^{-1}}}
\newcommand{\segue}{SEGUE}
\newcommand{\sdss}{SDSS}

\newcommand{\apoor}{\ensuremath{\alpha}-young}
\newcommand{\aenhanced}{\ensuremath{\alpha}-old}

\newcommand{\feh}{\ensuremath{[\mathrm{Fe/H}]}}
\newcommand{\afe}{\ensuremath{[\alpha\mathrm{/Fe}]}}

\newcommand{\Ro}{\ensuremath{R_0}}

\newcommand{\hs}{\ensuremath{R_\sigma}}

\newcommand{\rsun}{\ensuremath{R_{\odot}}}

\newcommand{\zmedian}{\ensuremath{z_{1/2}}}

\newcommand{\ba}{B12a}

\begin{document}

\submitted{Astrophys.~J.~{\bf 755}, 115}

\title{The vertical motions of mono-abundance sub-populations in the
  Milky Way disk}
\author{Jo~Bovy\altaffilmark{1,2,3},
  Hans-Walter~Rix\altaffilmark{4},
  David~W.~Hogg\altaffilmark{4,5},
  Timothy~C.~Beers\altaffilmark{6,7}, 
  Young Sun Lee\altaffilmark{7}, and
  Lan~Zhang\altaffilmark{4}}
\altaffiltext{\theaddress}{\label{IAS}\stepcounter{address} Institute
  for Advanced Study, Einstein Drive, Princeton, NJ 08540, USA}
\altaffiltext{\theaddress}{\label{Hubble}\stepcounter{address} Hubble
  fellow}
\altaffiltext{\theaddress}{\label{email}\stepcounter{address}
  Correspondence should be addressed to bovy@ias.edu~.}
\altaffiltext{\theaddress}{\label{MPIA}\stepcounter{address}
  Max-Planck-Institut f\"ur Astronomie, K\"onigstuhl 17, D-69117
  Heidelberg, Germany}
\altaffiltext{\theaddress}{\label{NYU}\stepcounter{address} Center for
  Cosmology and Particle Physics, Department of Physics, New York
  University, 4 Washington Place, New York, NY 10003, USA}
\altaffiltext{\theaddress}{\label{NOAO}\stepcounter{address} National
  Optical Astronomy Observatory, Tucson, AZ 85719, USA}
\altaffiltext{\theaddress}{\label{Michigan}\stepcounter{address}
  Department of Physics \& Astronomy and JINA (Joint Institute for
  Nuclear Astrophysics), Michigan State University, East Lansing, MI
  48824, USA}

\begin{abstract} 
  We present the vertical kinematics of stars in the Milky Way's
  stellar disk inferred from \sdss/\segue\ G-dwarf data, deriving the
  vertical velocity dispersion, $\sigma_z$, as a function of vertical
  height $|z|$ and Galactocentric radius $R$ for a set of
  `mono-abundance' sub-populations of stars with very similar
  elemental abundances $\afe$ and $\feh$. We find that all
  mono-abundance components exhibit nearly isothermal kinematics in
  $|z|$, and a slow outward decrease of the vertical velocity
  dispersion: $\sigma_z\bigl (z,R\,|\, \afe,\feh \bigr )\approx
  \sigma_z\bigl (\afe,\feh \bigr )\times\exp\bigl (-(R-R_0)/7\,
  \mathrm{ kpc}\bigr )$. The characteristic velocity dispersions of
  these components vary from $\sim$ 15 km s\inv\ for chemically young,
  metal-rich stars with solar $\afe$, to $\gtrsim 50$~km s\inv\ for
  metal-poor stars that are strongly $\afe$-enhanced, and hence
  presumably very old. The mean $\sigma_z$ gradient $(\dd \sigma_z /
  \dd z)$ away from the mid-plane is only $0.3\pm
  0.2$~km~s\inv~kpc\inv. This kinematic simplicity of the
  mono-abundance components mirrors their geometric simplicity; we
  have recently found their density distribution to be simple
  exponentials in both the $z$ and $R$ directions. We find a continuum
  of vertical kinetic temperatures ($\propto\sigma^2_z$) as a function
  of $\bigl (\afe,\feh \bigr )$, which contribute to the total stellar
  surface-mass density approximately as
  $\Sigma_{R_0}(\sigma^2_z)\propto \exp(-\sigma^2_z)$. This and the
  existence of isothermal mono-abundance populations with intermediate
  dispersions (30 to 40 km s\inv) reject the notion of a thin--thick
  disk dichotomy. This continuum of disk components, ranging from old,
  `hot', and centrally concentrated ones to younger, cooler, and
  radially extended ones, argues against models where the thicker disk
  portions arise from massive satellite infall or heating; scenarios
  where either the oldest disk portion was born hot, or where internal
  evolution plays a major role, seem the most viable. In addition, the
  wide range of $\sigma_z\bigl (\afe,\feh \bigr )$ combined with a
  constant $\sigma_z(z)$ for each abundance bin provides an
  independent check on the precision of the \segue-derived abundances:
  $\delta_{\afe}\approx 0.07$ dex and $\delta_{\feh}\approx 0.15$ dex.
  The slow radial decline of the vertical dispersion presumably
  reflects the decrease in disk surface-mass density. This measurement
  constitutes a first step toward a purely dynamical estimate of the
  mass profile of the stellar and gaseous disk in our Galaxy.
\end{abstract}


\section{Introduction}
The spatial structure and kinematics of stars in the Milky Way's disk
appear to be complex, varying distinctly with the elemental abundances
of different sub-populations: more metal-rich populations form
thinner, kinematically cooler sub-components, while metal-poor and
\afe-enhanced populations form thicker, kinematically hotter disk
components
\citep[\eg,][]{Fuhrmann98a,Chiba00a,Feltzing03a,Lee11b,Bovy12a}. The
elemental abundances of stars, in particular \feh\ and \afe,
encode---albeit in complex ways---information about their age and
birth location within the Galaxy \citep[\eg,][]{Schoenrich08a}.  The
present chemo-dynamical structure of the Milky Way's stellar disk must
reflect the combination of the orbits on which stars were formed, and
the evolutionary changes that they underwent since, with the
abundances being the best practical sub-population `tag' that is
preserved throughout the lifetime of a star. As such, the
chemo-dynamical structure of our Galaxy's disk holds unique clues
toward understanding its formation and evolution.  Because a good
portion of stars in the present-day universe live in galaxies
comparable to the Milky Way, and because the large majority of Milky
Way stars live in the disk, dissecting and understanding our Galactic
disk has broad implications for our understanding of galaxy formation.

In a recent paper, we have expanded on earlier work studying the
Galactic disk as a function of elemental abundances, by showing that
the geometric structure of the disk is relatively simple when viewed
as a superposition of `mono-abundance' sub-populations \citep[][\ba\
hereafter]{Bovy12a} drawn from \sdss/\segue\
\citep{Abazajian09a,Yanny09a}: the density of any set of stars chosen
to have a narrow range in $\afe$ and $\feh$ appears to be
well-described by a single exponential in both the radial and vertical
directions. In particular, \ba\ found that the (single) vertical scale
height, $h_z$ at a given $\bigl (\afe , \feh \bigr )$ varies
systematically with these two abundance parameters, in the sense that
populations that are metal poor and $\alpha$-enhanced, or \aenhanced,
are vertically thicker. It had been established that the thicker disk
components are \aenhanced\ \citep{Fuhrmann98a,Prochaska00a,Bensby03a},
but our recent work shows that the reverse also holds: \aenhanced\
sub-populations form thicker disk components than \apoor\
sub-populations. Furthermore, an analysis of the total amount of
stellar mass in disk components of different thicknesses exhibits a
continuous and monotonic distribution of disk thicknesses, rather than
a simple thin--thick dichotomy \citep{Bovy12b}.

In this present paper, we follow up on \ba\ by exploring the vertical
kinematics, $p(v_z | z,R)$ or $\sigma_z(z,R)$, for mono-abundance
sub-populations, in practice for subsets of G-type dwarfs from
\sdss/\segue\ in a small range of $\bigl (\afe , \feh \bigr )$.  The
goals of this analysis are two-fold: first, to see how the distinct
and simple spatial structure of mono-abundance sub-populations is
reflected in the vertical motions of these sub-components; second,
to lay the ground work for a dynamical analysis to determine the
gravitational potential near the disk plane.

Beyond the immediate solar neighborhood ($\sim\,100-200$ pc, \eg,
\citealt{Flynn94a}, \citealt{Nordstroem04a}), the first determination
of the vertical velocity dispersion profile was carried out by
\citet{Kuijken89a}, who obtained nearly complete line-of-sight
velocities for a sample of about 500 K-type dwarfs near the south
Galactic pole. For this sample, with a fairly broad metallicity
selection function, they found that the velocity dispersion profile,
$\sigma_z(z)$, rose from $\sim 18$~km s\inv\ near the mid-plane to
$\sim 40$~km s\inv\ at $\sim 1.3$~kpc above the
plane. \citet{Fuchs09a} used proper motions of M-type stars from
\sdss\ to confirm that the velocity dispersion exhibits a quite strong
linear increase away from the disk mid-plane, finding an even stronger
gradient. \citet{Bond10a} used photometric metallicity estimates to
isolate a low-metallicity sample, which showed a considerably slower
rise of $\sigma_z(z)$ away from the plane.

Much of the rise in $\sigma_z(z)$ found in these analyses is
presumably attributable to a much higher fraction of low-metallicity
stars well above the mid-plane, stars that form a kinematically
hotter sub-component. This would also explain why the sample with a
more restricted metallicity range \citep{Bond10a} shows a shallower
rise toward large $|z|$. Just recently, \citet{Liu12a} confirmed
this interpretation, using preliminary $\afe$ values for \sdss/\segue\
G-dwarf spectra. They show that the vertical dispersion profile
becomes approximately isothermal, when splitting the sample in a
number of $\afe$ bins.  In this paper, we expand on these studies, by
determining $\sigma_z(z)$ for sub-populations in a narrow range in
$\bigl (\afe , \feh \bigr )$, which turn out to be exquisitely close
to isothermal, \ie, $\sigma_z\bigl (z | R, \afe , \feh \bigr
)\approx$~constant.

When combined with the spatial structure in the radial and vertical
direction, the kinematics of these mono-abundance populations form the
basis for exploring the viability of various internal or external disk
evolution and disk heating mechanisms. The vertical density
distribution and kinematics of these mono-abundance populations can
also provide powerful constraints on the gravitational potential
perpendicular to the plane, for two reasons. First, components with a
simple spatial and kinematic distribution function are more easily
modeled, and a density profile $\nu_*(z |\afe , \feh )\propto
\exp{\bigr (-z/h_z(\afe , \feh )\bigr )}$, $\sigma_z(z| \afe , \feh
)\approx$~constant is about as simple as it could get.  Second, all
mono-abundance sub-components feel the same gravitational potential,
and hence provide extensive mutual checks on the dynamical inferences.

The outline of this paper is as follows. In
\sectionname~\ref{sec:data} we briefly summarize the properties of the
data set, and in \sectionname~\ref{sec:analysis} we describe the
approach to estimating $\sigma_z(z| \afe , \feh )$. We then present
the results of this analysis in \sectionname~\ref{sec:results}, not
only $\sigma_z(z|\rsun , \afe , \feh )$ but also its dependence on
Galactocentric radius. We also describe how our observations that the
mono-abundance components are so nearly isothermal in the vertical
direction and have such a strong dependence of the dispersion on
$\bigl (\afe , \feh \bigr )$ can be used to assess the abundance
errors of the \segue\ data, independent of all other existing
constraints.  In \sectionname~\ref{sec:discussion} we put these
results into context and discuss their implications. We summarize our
findings in \sectionname~\ref{sec:conclusion}.

\section{\sdss/\segue\ G-dwarf data}\label{sec:data}

\begin{figure*}[htbp]
\includegraphics[width=0.5\textwidth]{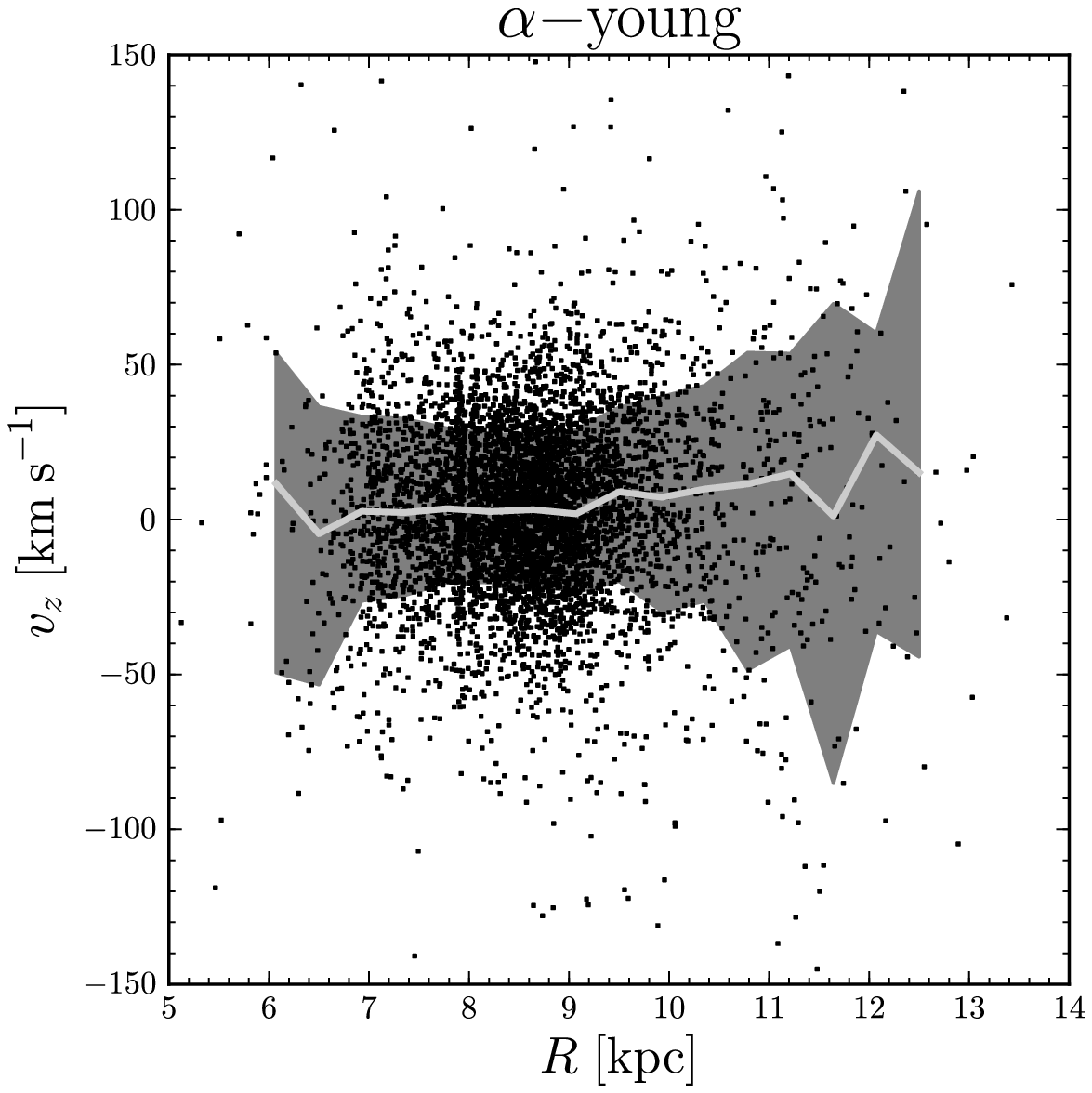}
\includegraphics[width=0.5\textwidth]{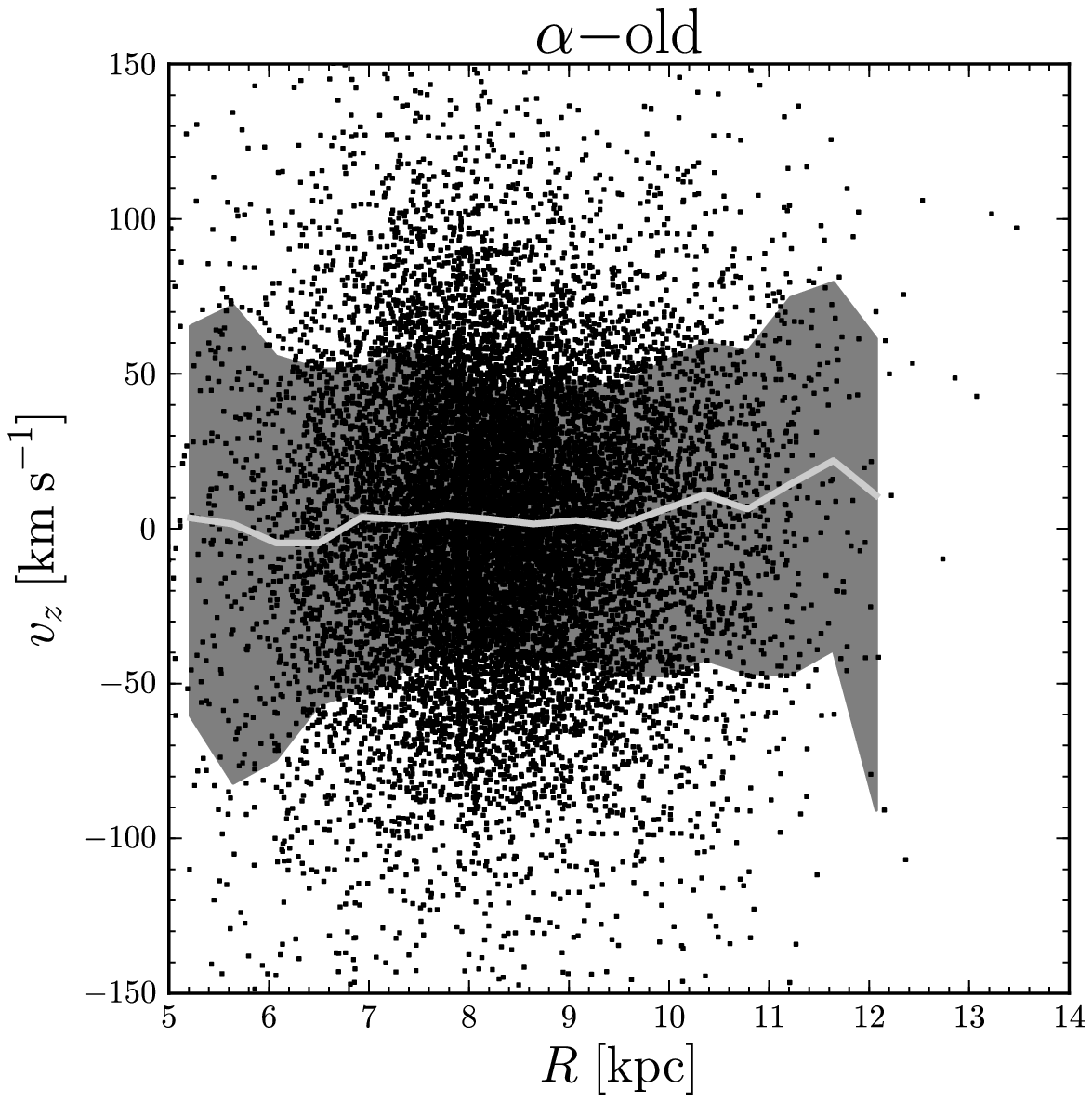}\\
\includegraphics[width=0.5\textwidth]{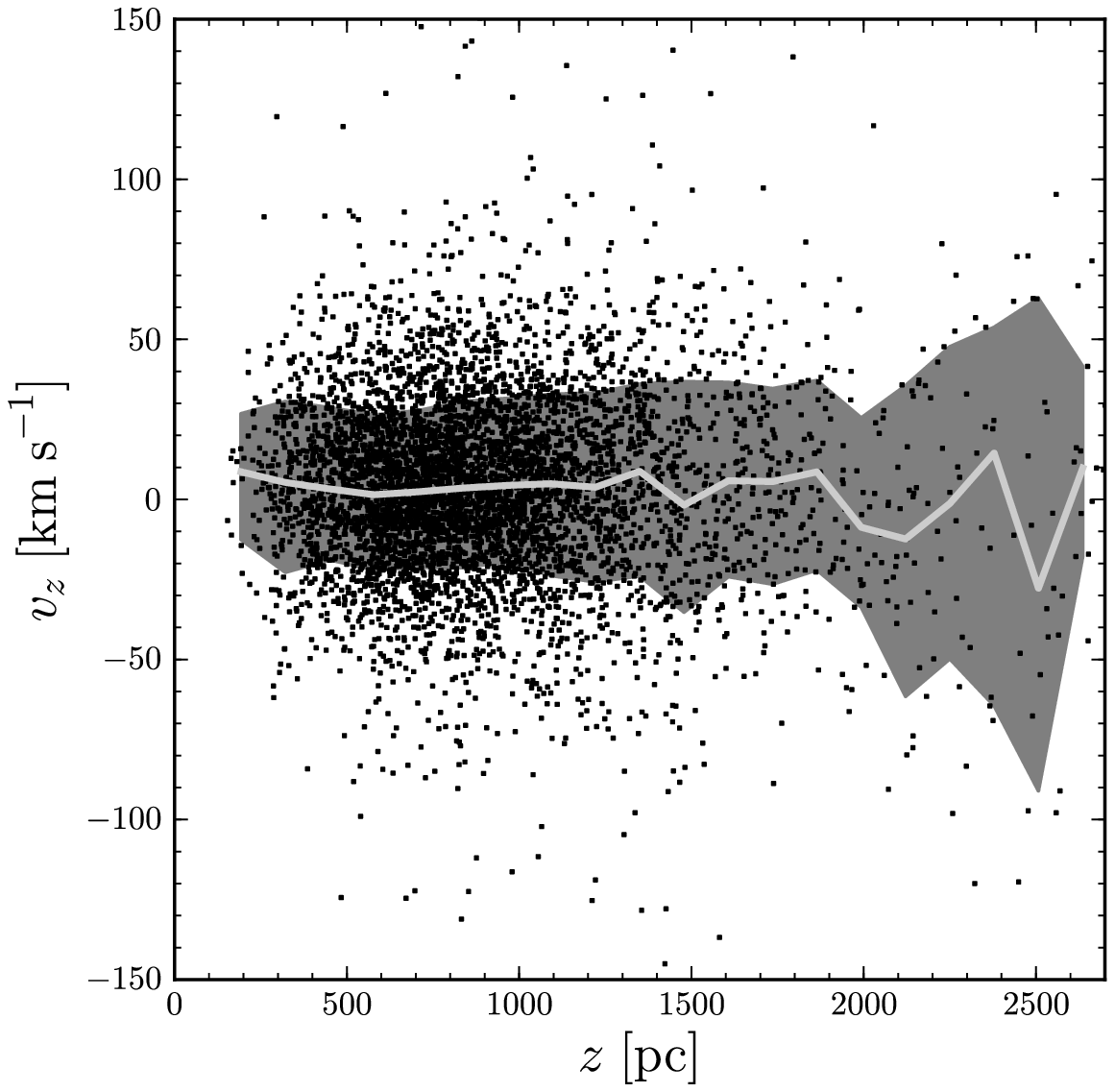}
\includegraphics[width=0.5\textwidth]{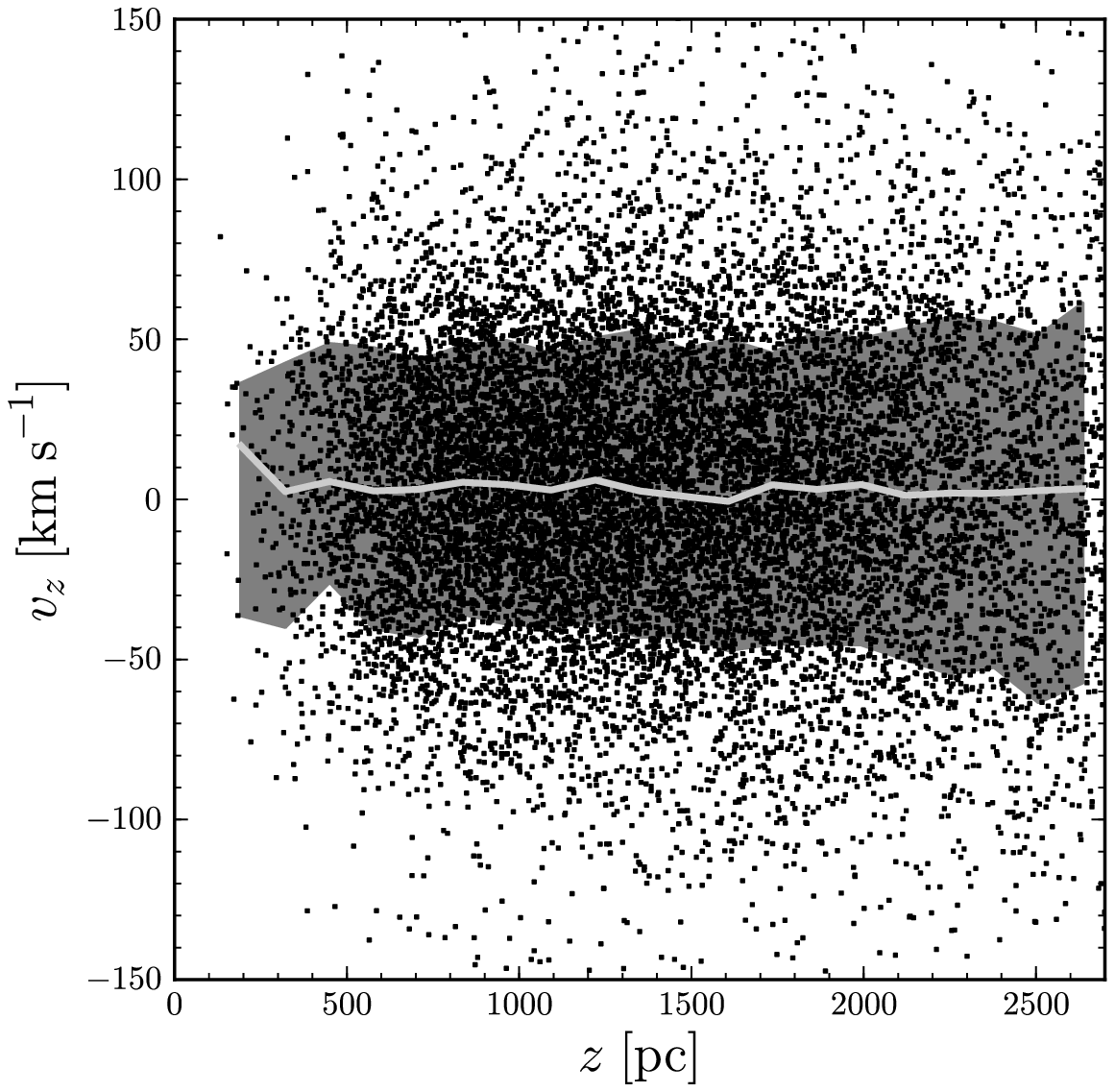}
\caption{Vertical velocities as a function of Galactocentric radius
  (top panels) and vertical height (bottom panels) for the \apoor\
  sample (left panels; $-0.3 < \feh < 0.25, 0.00 < \afe < 0.25$ ) and
  for the \aenhanced\ (right panels; $-1.5 < \feh < -0.25, 0.25 < \afe
  < 0.50$) sample.  The median and 16/84 quantiles of the distribution
  in bins of width $\approx 0.4$ kpc (as a function of $R$) and
  $\approx 150$ pc (as a function of $z$) are indicated by the solid
  line and the shaded regions, respectively.}\label{fig:datavz}
\end{figure*}

The sample on which we draw in this analysis is identical to the one
used in \ba, but we now also use the stars' line-of-sight velocities
and proper motions explicitly.  For an extensive description of the
sample, we refer the reader to \ba, \citet{Yanny09a}, and
\citet{Lee11b}, and just summarize some salient points here: G-type
dwarfs are the most extensive of the systematically targeted
sub-samples in \segue\ to explore the Galactic disk; they are the most
luminous tracers whose main-sequence lifetime is larger than the
expected disk age at basically all metallicities.  Their rich
metal-line spectrum affords good velocity determinations ($\sim\!  5$
to 10~km s\inv; \citealt{Yanny09a}), as well as good abundance ($\afe
, \feh$) determinations ($\delta_{\afe}\sim 0.1$ dex, $\delta_{\feh}\sim
0.2$ dex\footnote{In this paper, we use $\delta$ to indicate
  observational uncertainties, to avoid confusion with the dispersion
  symbol $\sigma$.},
\citealt{Lee08a,Lee08b,AllendePrieto08a,Schlesinger10a,Lee11a,Smolinski11a},
which we re-examine in this paper). Given the (metallicity-dependent)
absolute magnitude of targeted G-type dwarfs with $0.48<(g-r)_0<0.55$,
ranging from $4 \lesssim M_r \lesssim 6$, the distances to the sample
stars range from $0.6$~kpc to nearly 4~kpc. Stars somewhat closer to
the disk plane are sampled by the lines of sight at lower Galactic
latitudes, but the effective minimal distance limit of the stars
(600~pc) implies that the vertical heights below one scale height
($|z| < h_z$) of the thinner disk components is not sampled by the
data. As in \ba, we employ a signal-to-noise ratio cut of S/N $> 15$,
rather than the $S/N > 20$ recommended by \citet{Lee11a}; we have
explicitly checked that our results are the same when using S/N $>
20$.

The specific analysis in this paper requires the height above the
plane, the Galactocentric radius, and the vertical velocity and its
error for each star. For a star at distance $d$ in the direction
($l,b$), \ie, with $\vec{d}=d\cdot (\cos l \cos b , \sin l \cos b,
\sin b)$ in the Cartesian Galactic coordinate system, we have $|z| =
|\vec{d}\cdot \vec{e_z}|$ and $R = |\vec{R_0}+\vec{d}-\vec{z}|$. We
assume that the Sun's displacement from the mid-plane is 25 pc toward
the north Galactic pole \citep{Chen01a,Juric08a}, and that the Sun is
located at 8 kpc from the Galactic center \citep[\eg,][]{Bovy09b}. We
ignore distance uncertainties when using the spatial position of stars
in the sample in the analysis below, as the distance uncertainties are
much smaller than the spatial gradients in the quantities of
interest. We do, however, propagate the distance uncertainties into
the velocity uncertainties.

The motions in the $z$-direction are a combination of the
line-of-sight velocities, $v_{\mathrm{los}}$, and the proper motions in
the $b$-direction, $\mu_b$: $v_z (b,d,v_{\mathrm{los}}) =
v_{\mathrm{los}}\,\sin b + d\,\mu_b$; we assume that the Sun's
vertical motion is 7.25 km s\inv\ \citep{Schoenrich10a}.  Given that
we expect $\sigma_z\lesssim 20$~km s\inv\ for the coldest disk
components, the $v_z$ errors, $\delta_{v_z}$, matter greatly, both
because for stars of lower signal-to-noise ratio
$\delta_{v_{\mathrm{los}}}\approx 10$~km s\inv, and because with
typical $\delta_{\mu_b}\sim 3.5$~mas yr\inv\ \citep{Munn04a} the error
contribution from the proper motion $\delta_{v_z|\mu_b} =
15\,\mathrm{km\ s}\inv\cdot [d/1\,\mathrm{kpc}]\cdot [\delta_{\mu_b} /
(3.5\, \mathrm{mas\ yr}\inv)]$ rises linearly with distance, becoming
dominant beyond 1~kpc. The uncertainties in $v_z$ for each star are
derived by marginalizing over the line-of-sight velocity, proper
motion, and distance errors; the distance errors in turn are obtained
by marginalizing over the color and apparent-magnitude errors that
enter the photometric distance relation (Equation~(A7) of
\citealt{Ivezic08a}, see also \ba), which is assumed to have an
intrinsic scatter of 0.1 mag in the distance modulus.  It is important
to note that for the present analysis, the detailed understanding of
the sampling function that was painful and crucial for \ba\ and
\citet{Bovy12b} plays no role here, as long as one can assume that the
vertical velocities of the targeted G-type dwarfs at a given
$l,b,r,g-r,$ and \afe, \feh\ have kinematics identical to their
untargeted kin with those same non-kinematic parameters.

As discussed in detail in Section 5.1 of \ba, the distances we use
here could be systematically under- or over-estimated by up to 10\,\%
due to the presence of unresolved multiple systems or the use of a
different distance calibration, respectively. This systematic distance
uncertainty can propagate into the vertical velocities for stars in
our sample through the dependence on the distance of the velocity
component that is tangential to the line of sight. In what follows we
discuss the impact of this systematic distance error on our results,
finding that it does not significantly impact any of our results.

We begin with a \segue\ G-dwarf sample that has about 28,000 stars
with acceptably well-determined measurements, but here we only use
those 23,767 stars that fall within well-populated `mono-abundance'
bins in the (\feh,\afe) plane (\ba; a bin is well-populated when it
contains more than 100 stars; the maximum number of stars in a bin is
789). \figurename~\ref{fig:datavz} provides a basic representation of
the data used in the subsequent analysis: the top panels show $v_z$
\vs\ $R$ for a broadly selected \apoor\ (left; $-0.3 < \feh < 0.25,
0.00 < \afe < 0.25$) and \aenhanced\ (right; $-1.5 < \feh < -0.25,
0.25 < \afe < 0.50$) sample, respectively. The bottom panels show
$v_z$ \vs\ $z$.  This figure immediately illustrates that, as
expected, the vertical velocity dispersion of the \aenhanced\ sample
is considerably higher than that of the \apoor\ sample, and that there
are no strong changes of the velocity dispersion with either distance
from the plane or Galactocentric radius. The spatial distribution of
these two coarse-binned abundance-selected samples is shown in Figure
3 of \ba. That figure shows that the sample spans multiple kpc in $R$
for \apoor\ and \aenhanced\ subsamples and that all subsamples cover
the $(R,z)$ plane fairly well. Thus, we can model the full $(R,z)$
dependence of $\sigma_z$ in each mono-abundance bin.

\section{Mapping the vertical velocity dispersion of G-type dwarfs}\label{sec:analysis}

We now describe how we estimate the vertical velocity distribution
profile for each mono-abundance sub-population, characterized by
$\sigma_z(z,R | \afe , \feh )$, exploiting the discrete information we
have: $\{ v_{z,i}, \delta_{v_{z,i}} | z_i, R_i\}, i=1,N_j$ for each
abundance bin $(\feh_j,\afe_j)$. Note that we model the velocity
distribution as Gaussians, rather than directly measuring it as the
second moment of the observed vertical velocities
$\sum_i{(v_{z,i}-\langle v_z \rangle)^2/N}$ in spatial bins. This
allows us to take the individual observational uncertainties
$\delta_{v_{z,i}}$ into account and to deal with outliers. As the
\segue\ data selection is kinematically unbiased and we are only
interested in the distribution of vertical velocities as a function of
position, we do not have to correct for any selection effects.

We presume that the vertical velocity distribution $p(v_z | z,R,
\feh_j,\afe_j)$ can be described by
\begin{equation}
\begin{split}
p( v_z | z,R, \vec{p}_j, \delta_{v_{z,i}}\, )= & (1-\epsilon_{j})\cdot \frac{1}{\sqrt{2\pi}\sigma_z(z,R\, |\, \vec{p}_j,\delta_{v_{z,i}})}
\\ & \qquad \qquad \cdot \exp{\Bigl (\frac{-v^2_z}{2\sigma^2_z(z,R\, |\, \vec{p}_j,\delta_{v_{z,i}})}}\Bigr )\ \\ & + \epsilon_{j}\cdot p_{\mathrm{backgr}}(z,R\,),
\end{split}
\end{equation}
where $\vec{p}_j$ is the vector of parameters to be fit in each
abundance bin $j$ and $p_{\mathrm{backgr}}(z,R)$ is a simple normalized
interloper model (reflecting data analysis outliers and halo
contamination). Specifically, for $p_{\mathrm{backgr}}(z,R)$ we choose
a normalized Gaussian with a dispersion of 100~km s\inv, convolved
with the observational uncertainty $\delta_{v_{z,i}}$, and
$\epsilon_j$ is a free parameter that is the fraction of stars deemed
interlopers. In what follows we will fit $\vec{p}_j$ and $\epsilon_j$
for each abundance bin independently. To reduce notational clutter, we
will mostly drop the `$j$' index with the assumptions that the
parameters $\vec{p}$ and $\epsilon$ are fitted in the discrete
$(\feh_j,\afe_j)$ abundance bins.

\begin{figure*}[htbp]
\includegraphics[width=0.5\textwidth,clip=]{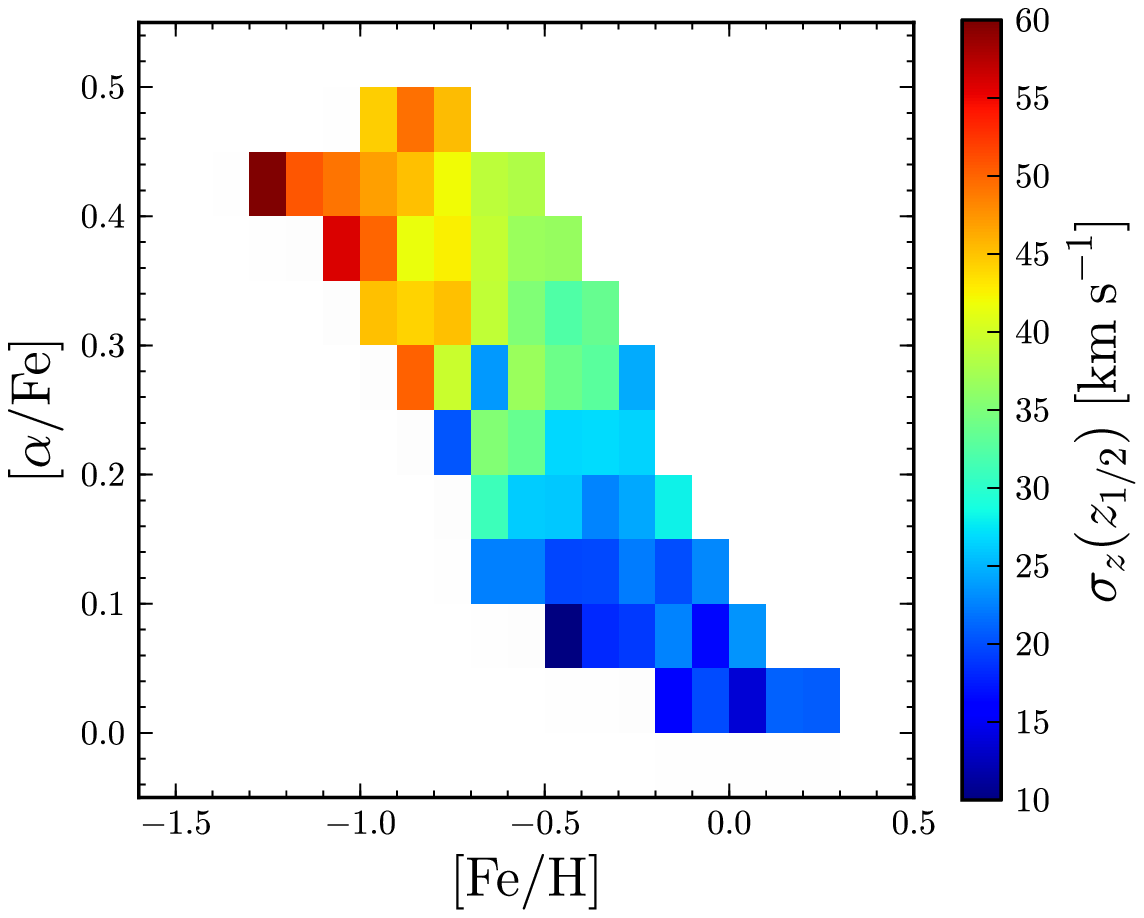}
\includegraphics[width=0.5\textwidth,clip=]{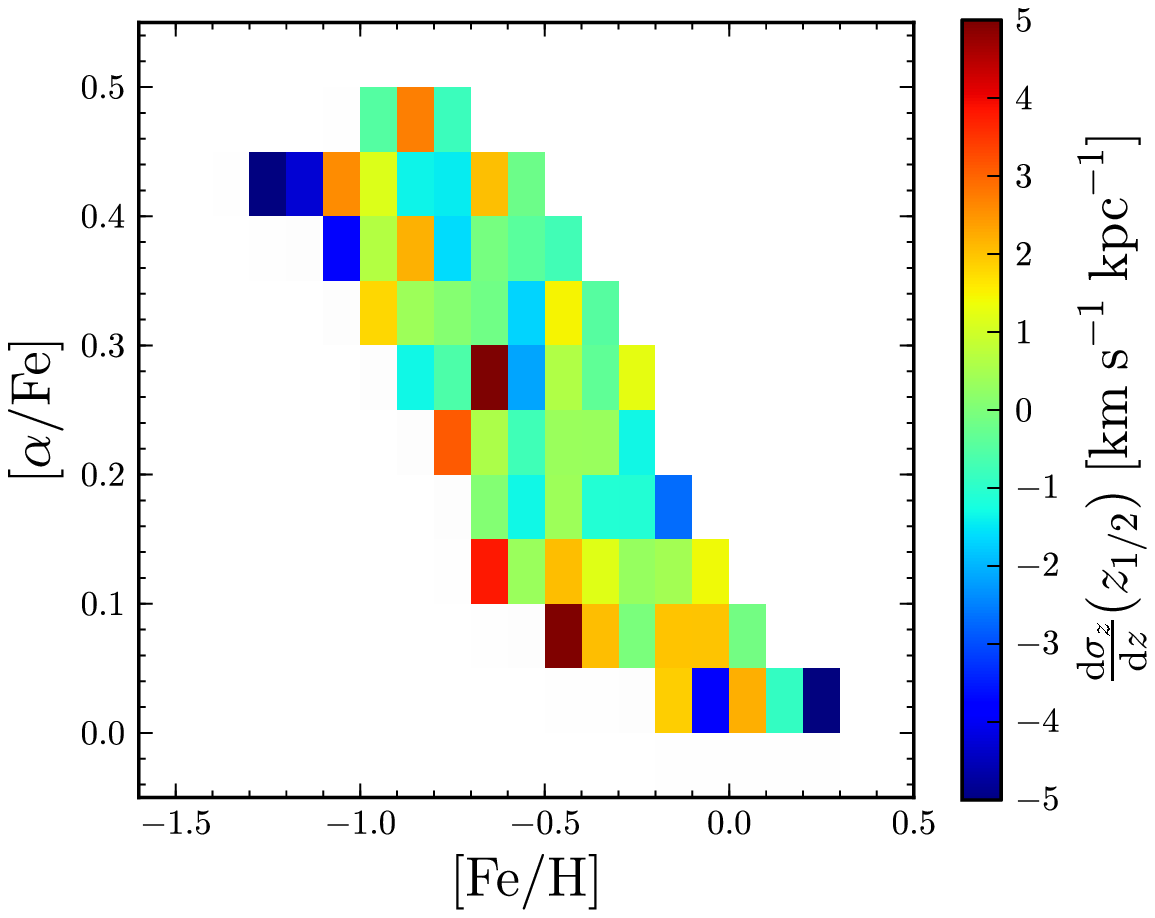}\\
\includegraphics[width=0.5\textwidth,clip=]{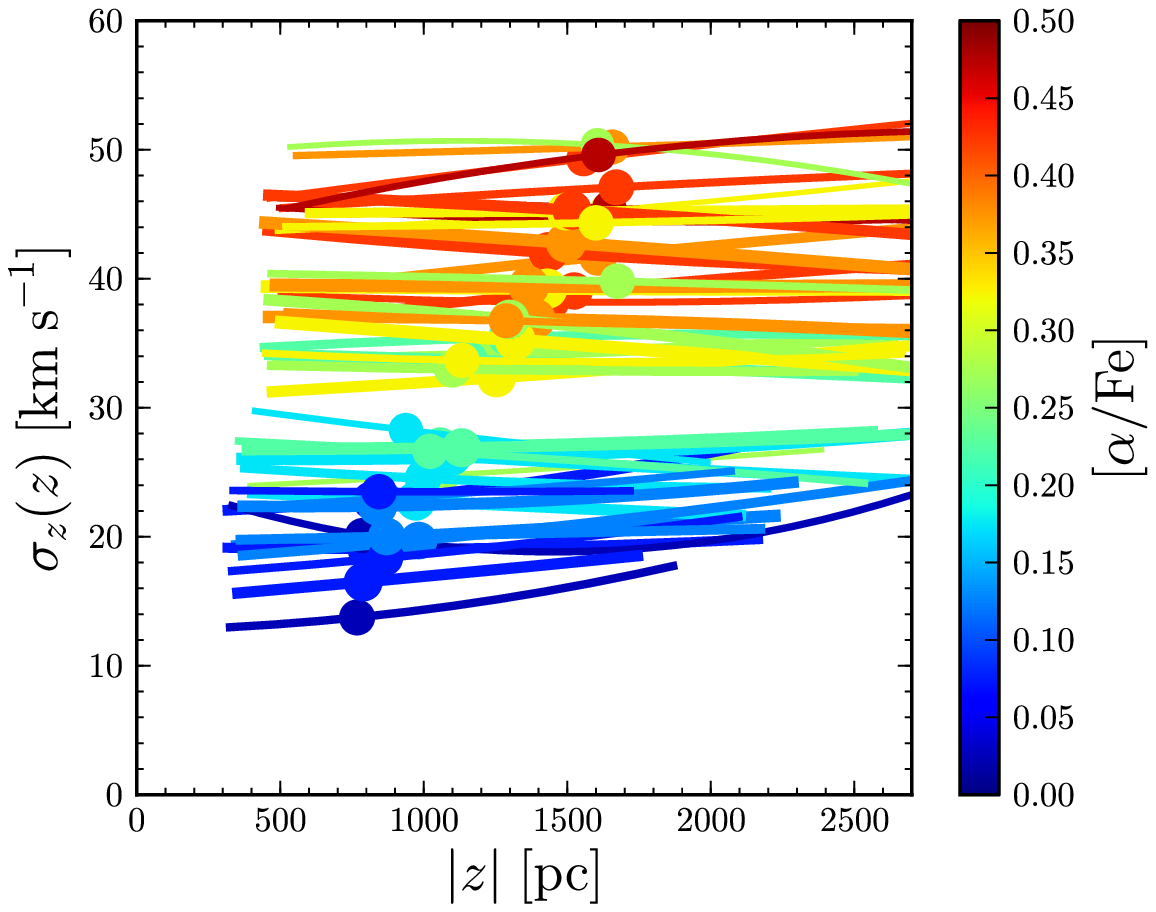}
\includegraphics[width=0.5\textwidth,clip=]{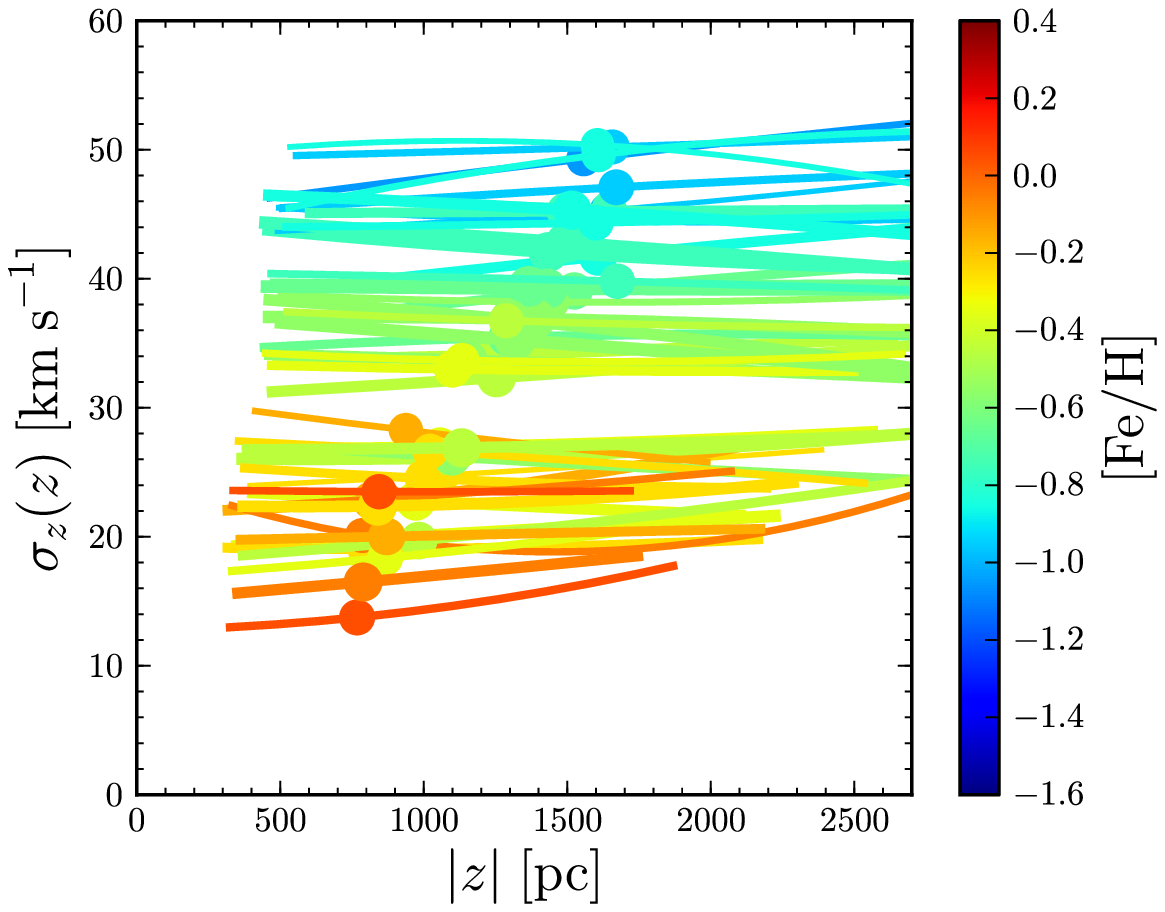}
\caption{Best-fit vertical velocity dispersion profile, $\sigma_z(z)$,
  at \Ro\ and slope of the velocity dispersion as a function of $|z|$,
  evaluated at the median vertical height \zmedian\ for each bin, as a
  function of \feh\ and \afe\ (top panels) in bins of width 0.1 in
  \feh\ and 0.05 in \afe. The bottom panels show the best-fit velocity
  dispersion profile (modeled as a 2nd-order polynomial) as a function
  of height above the mid-plane for each bin in the top panels with a
  20\,\% or better determination of $\sigma_z(\zmedian)$, color-coded
  for each point in (\feh,\afe), according to \afe\ (left panel) and
  \feh\ (right panel). The velocity dispersion in the bottom panels is
  evaluated over the range in vertical height that contains
  95\,\% of the data sample for each mono-abundance bin; the
  median height of the observed data is indicated by a dot in the
  bottom panels. Each (\feh,\afe) bin contains at least 100 stars, and
  the width of each line in the bottom panels is proportional to the
  square root of the number of data points.}\label{fig:results}
\end{figure*}

\begin{figure*}[htbp]
\includegraphics[width=0.5\textwidth]{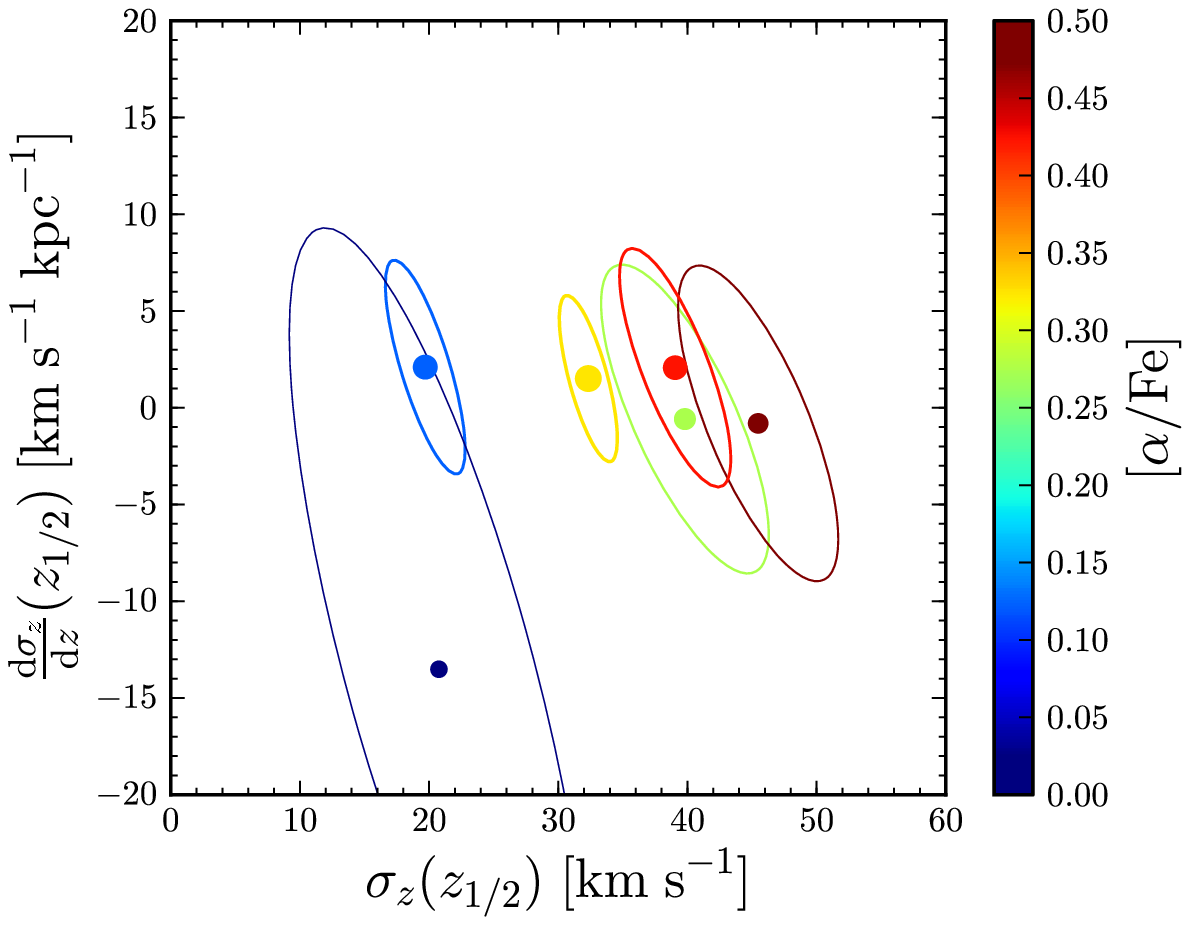}
\includegraphics[width=0.5\textwidth]{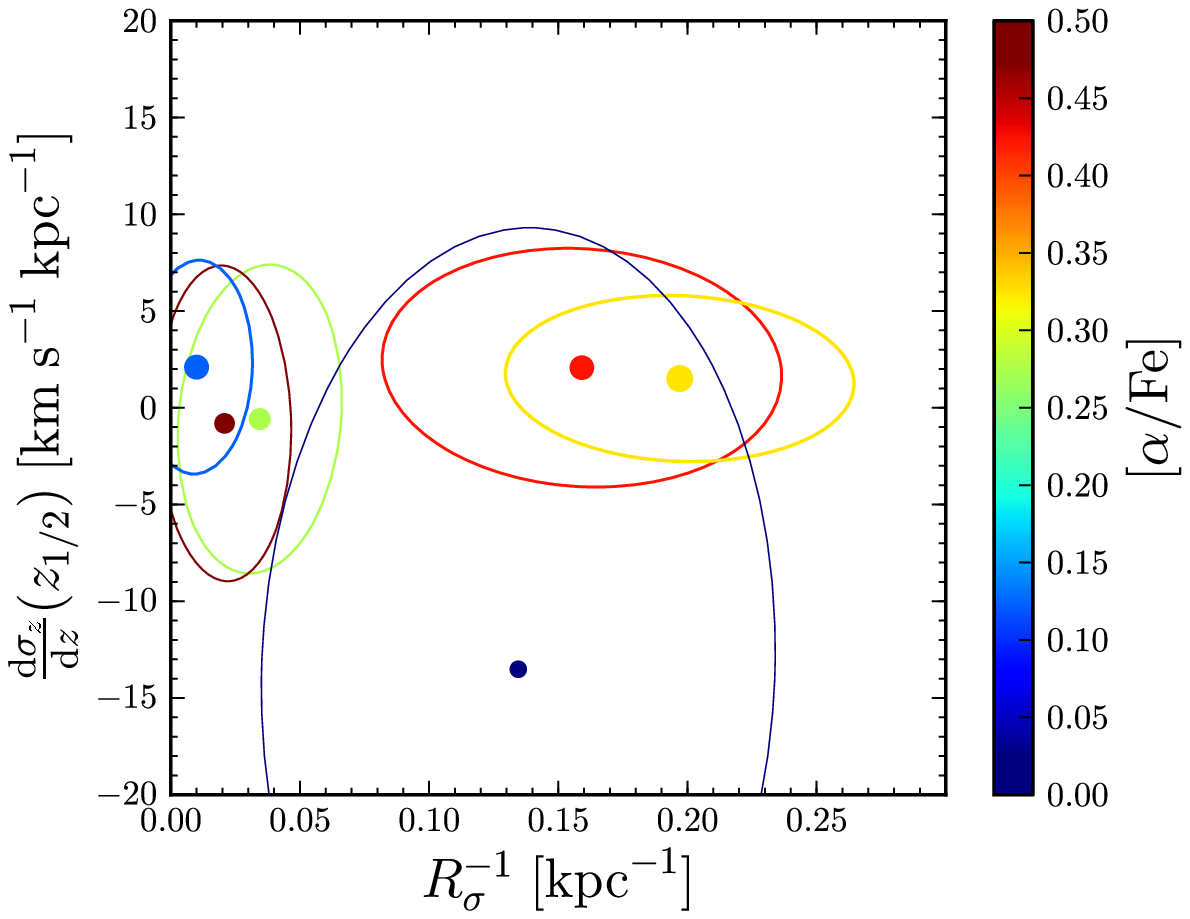}\\
\includegraphics[width=0.5\textwidth]{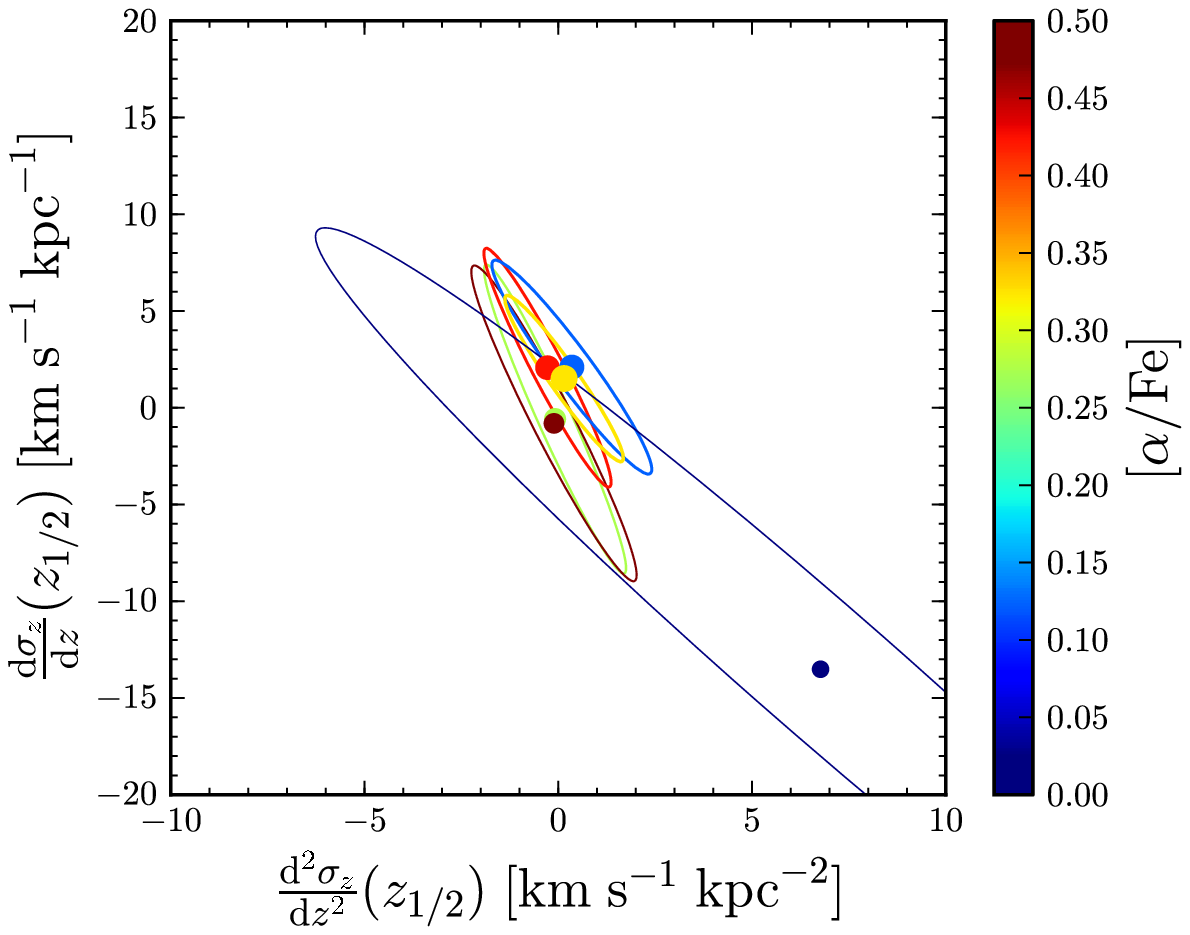}
\includegraphics[width=0.5\textwidth]{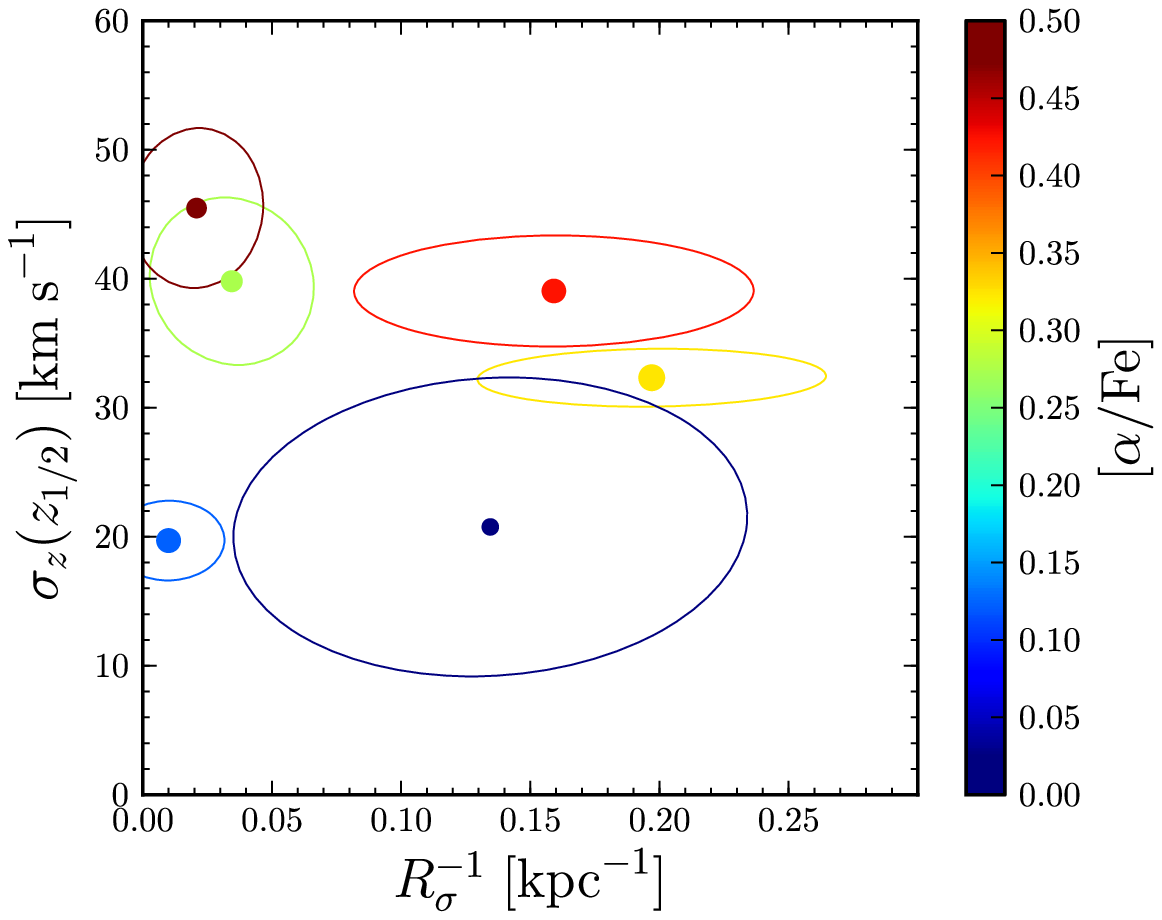}
\caption{Parameter uncertainties for the quadratic velocity dispersion
  profile fits for six representative mono-abundance bins: the peak
  and one-sigma uncertainty ellipses for each (\feh,\afe) bin's PDF
  for the parameters of the vertical velocity dispersion are shown for
  2D projections of the 5D parameter space. Clockwise from the top,
  left panel these are: slope \vs\ velocity dispersion, slope \vs\
  inverse radial dispersion scale length, velocity dispersion \vs\
  inverse radial scale length, and slope \vs\ quadratic term of the
  vertical velocity profile. There are no significant correlations
  between the parameters, except for a strong anti-correlation between
  the slope and the quadratic term of the vertical dependence of
  $\sigma_z(z)$.}\label{fig:slopehsm}
\end{figure*}

In each mono-abundance bin, the velocity distribution is presumed
Gaussian with zero mean and a dispersion $\sigma_z(z,R\, |\,
\vec{p})$, described by a quadratic function in $|z|$, as we have no
strong priors on the functional form of the vertical dispersion
profile, except that it be smooth.  We assume that the radial
dependence of the velocity dispersion separates from the vertical
dependence, and that it is exponential:
\begin{equation}\label{eq:radial}
\begin{split}
\sigma_z^2(z,R\, |\, \vec{p},\delta_{v_{z,i}}) = & \sigma^2_z(z,R_0\, |\, p_1,p_2,p_3)\\ & \qquad \times \exp\left [-\, 2\,p_4\cdot (R-R_0 )\right ] + \delta^2_{v_{z,i}}\,
\end{split}
\end{equation}
where $p_4$ is the inverse scale length \hs\inv\ for the radial change
in the vertical velocity dispersion. The parameters $p_1, p_2$, and
$p_3$ describe the vertical profile of the velocity dispersion, with
$p_1 \equiv \sigma_z(\zmedian)$, $p_2 \equiv \dd \sigma_z(\zmedian) /
\dd z$, and $p_j^3 \equiv \dd \sigma^2_z(\zmedian) / \dd z^2$;
$\zmedian$ is the median vertical height above the mid-plane for each
mono-abundance data sample. The addition of the observational
uncertainty $\delta^2_{v_{z,i}}$ is due to the convolution of the
model with the Gaussian uncertainty model. Assuming uninformative flat
priors on $\vec{p}$ for each mono-abundance sub-population, we then
derive, through ensemble-based MCMC sampling (\citealt{Goodman10a};
\citealt{FM12a}), the posterior probability distribution function
(PDF) for the parameters of the velocity dispersion,
$\vec{p}_j(\feh_j,\afe_j)$, and the contamination fraction,
$\epsilon_{j}(\feh_j,\afe_j)$, in each mono-abundance bin
$(\feh_j,\afe_j)$. After correcting for outliers, the chi-squared per
degree of freedom of the fits in each bin are approximately one to the
level expected for the number of data points, indicating that this
model for the vertical velocity distribution is a good model.

To cast these parameterized models into the more intuitive terms of
$\sigma_z(z,R | \afe , \feh )$, one can then use samples from the PDF
of $\vec{p}_j$ to obtain representations of plausible dispersion profiles,
\eg, $\sigma_z(z)$ at $R_0$ and some $(\feh,\afe)$. In each
mono-abundance bin the contamination fraction $\epsilon_{j}$ is less
than a few percent, and is not discussed further.

\section{Results}\label{sec:results}

\begin{figure}[htbp]
\includegraphics[width=0.5\textwidth]{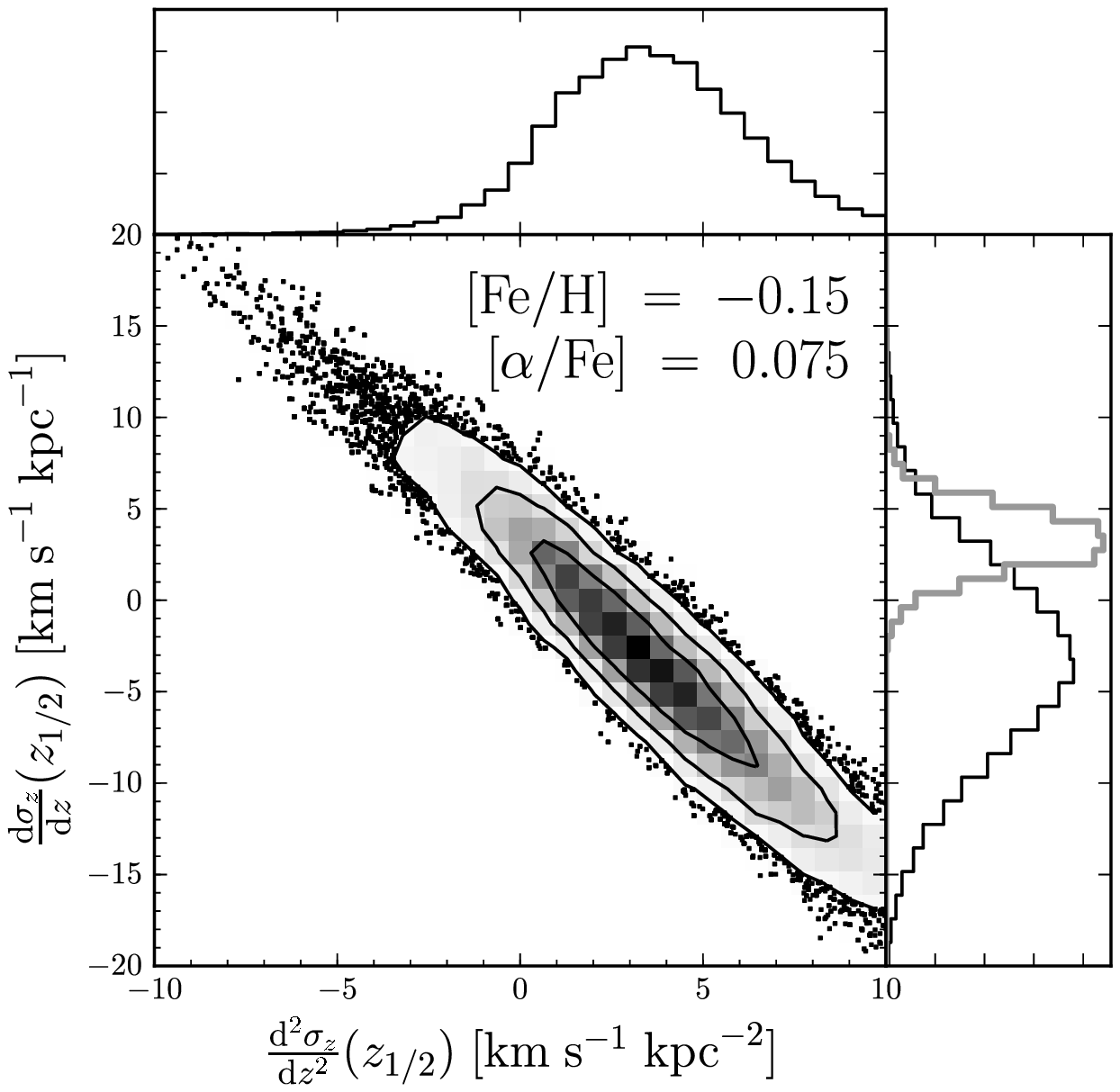}
\includegraphics[width=0.5\textwidth]{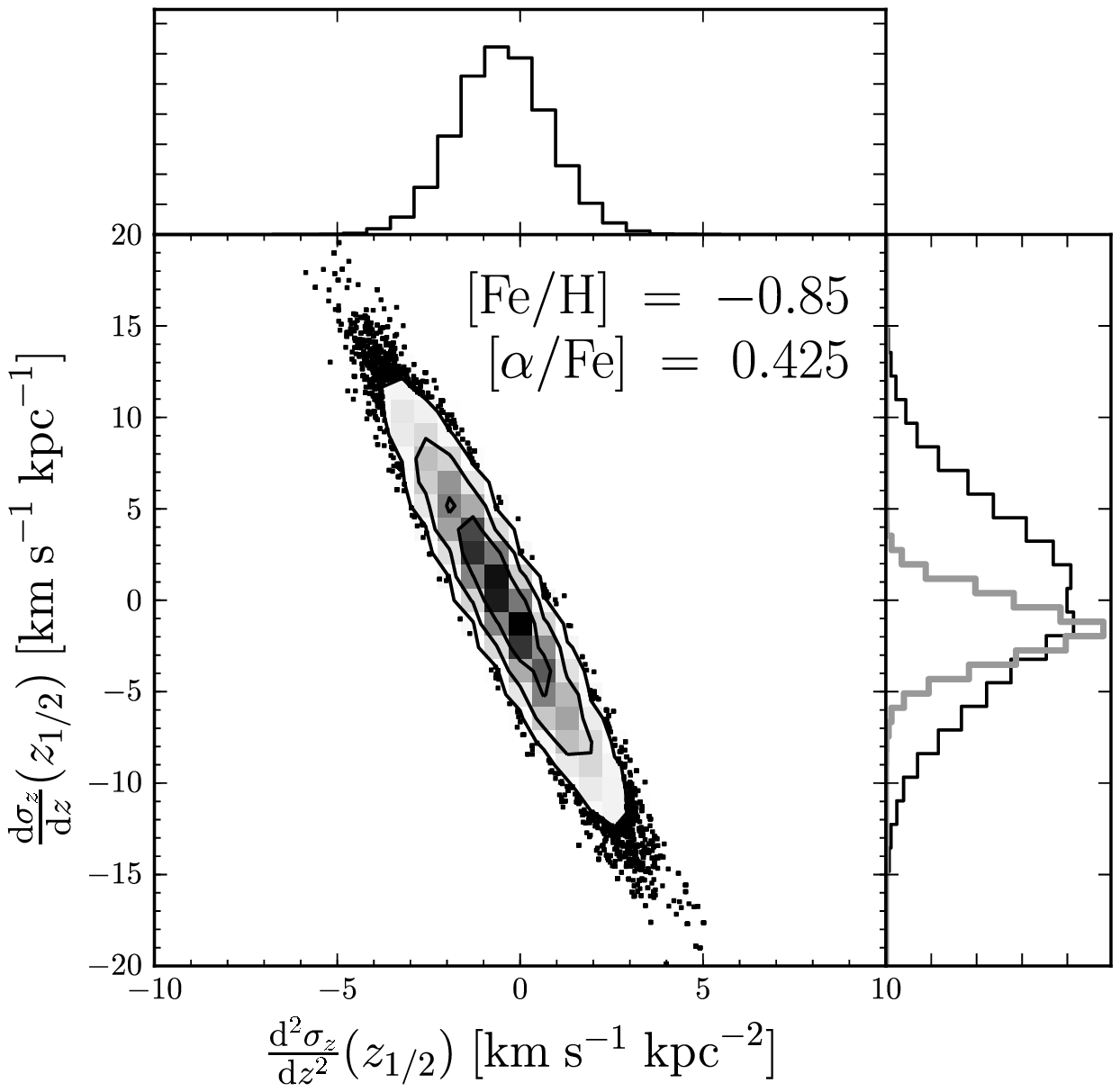}
\caption{Individual PDFs for the slope and quadratic term of the
  vertical-velocity-dispersion profile for two mono-abundance
  bins. The grayscale is linear and contours contain 68\,\%, 95\,\%,
  and 99\,\% of the distribution and points beyond the last contour
  are individually shown. For each mono-abundance bin, the slope and
  quadratic term of the vertical dependence of $\sigma_z(z)$ are
  strongly anti-correlated, in such a way as to minimize the total
  change in $\sigma_z$ with $z$. The gray histogram in the right inset
  of each figure shows the PDF for the slope assuming no quadratic
  term (\ie, just fitting a linear profile to the vertical dependence
  of $\sigma_z(z)$); in this case the slope is tightly constrained to
  be near zero.}\label{fig:indivPDFs}
\end{figure}

\subsection{Vertical dispersion profile: $\sigma_z(z,R_0\, |\,  \feh,\afe)$}\label{sec:vertical}

In \figurename s~\ref{fig:results}--\ref{fig:indivPDFs} we present the
results of our procedure for mapping the vertical velocity dispersion
of the \segue\ G-dwarf sample, divided in narrow bins in
(\feh,\afe). As described above, this procedure is robust to outliers
and automatically handles the individual uncertainties in the observed
vertical velocities. \figurename~\ref{fig:results} shows the main
result of this paper, namely, that the vertical velocity dispersion
profiles of mono-abundance populations, $\sigma_z(z,R_0\, |\, \afe ,
\feh )$, are nearly constant with height $z$ at some
$\sigma_z(\feh,\afe)$---\ie, they are isothermal---and that
$\sigma_z(\feh,\afe )$ varies distinctly across chemically-selected
sub-populations among the different abundance bins.  The two bottom
panels of \figurename~\ref{fig:results} show the vertical dispersion
profiles, $\sigma_z(z,R_0\, |\, \vec{p}_{best} )$, for each abundance
bin with a 20\,\% or better determination of $\sigma_z(\zmedian)$,
where $\vec{p}_{best}$ represents the peak of the PDF for the model
parameters, and the color coding reflects \afe\ (left) and \feh\
(right), respectively. The lines are drawn over the central 95\,\% of
the $|z|$-range of stars in each bin, reflecting the
abundance-dependent distance distribution in the sample.  Remarkably,
$\sigma_z$ is nearly constant (within a few km s\inv) for all
individual $(\feh,\afe )$ bins, while the $\sigma_z$ values vary from
15~km s\inv\ to 50~km s\inv\ among the bins. That variation in
$\sigma_z(\zmedian,R_0\, |\, \vec{p}_{best}, \feh,\afe)$, where
$\zmedian$ is the median vertical height above the mid-plane for each
mono-abundance data sample, is illustrated in the top left panel of
\figurename~\ref{fig:results}, showing that the vertical velocity
dispersion increases toward higher $\afe$ and lower $\feh$ in a
pattern that is qualitatively very similar to the corresponding
scale-height map of mono-abundance components, $h_z(\feh,\afe)$, from
\ba. The top right panel shows the linear slope of the
$\sigma_z(z,R_0\, |\, \feh,\afe )$ profile, ${\mathrm
  d}\sigma_z(z)/{\mathrm d}z$ evaluated at \zmedian.  The slopes are
typically at the level of 1 to 3 km s\inv\ kpc\inv, with no
recognizable dependence on $\afe$ and $\feh$. We return to this
remarkable degree of isothermality below.

While \figurename~\ref{fig:results} summarizes the best-fit
$\sigma_z(z,R_0\, |\, \feh,\afe )$, \figurename~\ref{fig:slopehsm}
illustrates the uncertainties in the inferred parameters $\vec{p}$ for
seven representative mono-abundance bins (the uncertainties for the
other bins are similar). The PDFs are summarized by the peak and
1-sigma uncertainty ellipses of two-dimensional projections of the
full five-dimensional parameter space for each mono-abundance
bin. This figure shows that the uncertainty on $\sigma_z$ is typically
a few km s\inv, such that the slight gap apparent in
\figurename~\ref{fig:results} around $\sigma_z = 30$ km s\inv\ is not
significant. This level of uncertainty also does not affect the large
abundance-trend of $\sigma_z(\zmedian,R_0)$ between 10 and 60 km
s$^{-1}$ in the top panel of
\figurename~\ref{fig:results}. \figurename~\ref{fig:slopehsm} also
shows that there are no significant correlations among the parameters,
except for a slight correlation between the slope and the magnitude of
the dispersion and a strong correlation between the slope and
quadratic term of the vertical dependence of the velocity
dispersion. The slope and quadratic term are anti-correlated, such
that the overall change in velocity dispersion over the $|z|$-range of
the sample is small in all cases. The parameters of the vertical
dependence of $\sigma_z$ are not correlated with the inverse scale
length \hs\inv; we discuss the inferred inverse scale length
further below. \figurename~\ref{fig:indivPDFs} shows the full PDF for
the slope and the quadratic term of the vertical dependence of $\sigma_z$
for two mono-abundance bins. These full PDFs further illustrate the
strong anti-correlation between the slope and quadratic term. This
strong anti-correlation does not mean that the data only constrain the
vertical slope of the vertical velocity dispersion profile without
constraining the second derivative. The fits strongly prefer a flat
vertical profile for the velocity dispersion over the full vertical
range of $\approx 2$ kpc of the data in each mono-abundance bin.

By combining the individual mono-abundance PDFs for the slope and the
quadratic term we obtain a joint estimate for the vertical dependence
of the vertical velocity dispersion. The peak of the joint PDF occurs
at ${\mathrm d}\sigma_z(z)/{\mathrm d}z = 0.7 \pm 0.5$ km~s\inv\
kpc\inv, ${\mathrm d}^2\sigma_z(z)/{\mathrm d}z^2 = -0.12 \pm 0.2$
km~s\inv\ kpc$^{-2}$, with strongly correlated error bars. Assuming no
quadratic term, the joint estimate for the slope is ${\mathrm
  d}\sigma_z(z)/{\mathrm d}z = 0.3 \pm 0.2$ km~s\inv\ kpc\inv. Thus,
each individual mono-abundance bin is consistent with isothermality at
the $\sim\!10$\,\% (few km s\inv) level, and the mean slope is
consistent with zero at the percent level (few tenths of a km s\inv),
when compared with the range in vertical velocity dispersion present
in the full disk sample.

It is hard to judge from \figurename~\ref{fig:slopehsm} whether the
individual slope and quadratic term estimates and uncertainties are
consistent with the joint estimate, in the sense of these individual
estimates really being noisy estimates of the joint estimate. To
answer this question, we calculate for each individual mono-abundance
(slope and quadratic coefficient) PDF the quantile of the distribution
at which the joint estimate lies. In practice, this calculation is
done by binning the individual PDF in two dimensions, sorting the
binned PDF in ascending order (by flattening the 2D binned PDF into a
1D list), and cumulatively summing the resulting list and normalizing
the total sum to one. This list then contains the quantile of the PDF
at which each bin in the binned PDF lies, and we can find the bin and
quantile corresponding to the jointly estimated slope and quadratic
coefficient. We find that the distribution of quantiles thus
calculated is relatively flat, with no noticeable dependence on \afe,
indicating that the individual (slope, quadratic coefficient)
estimates and uncertainties are consistent with being noisy estimates
of the joint slope and quadratic coefficient.

As discussed in \sectionname~\ref{sec:data}, the distances used could
be systematically under- or over-estimated by up to 10\,\%. Through
the dependence on the distance of the velocity component tangential to
the line of sight this could lead to a increase or decrease in the
velocity dispersion's vertical gradient. To investigate this we have
repeated the analysis while systematically changing the distances by
10\,\%. We find for both under- and over-estimated distances that the
joint estimate for the slope remains consistent with zero, with
$\mathrm{d}\sigma_z(z)/{\mathrm d}z = 0.4 \pm 0.2$ km~s\inv\ kpc\inv\
and $\mathrm{d}\sigma_z(z)/{\mathrm d}z = 1.1 \pm 0.4$ km~s\inv\
kpc\inv, for distances that are 10\,\% smaller and larger,
respectively. Additionally, we have repeated the analysis considering
only stars with $|b| > 50^\circ$, for which most of the vertical
velocity comes from the line-of-sight velocity measurement, which is
not affected by distance systematics---and a flat vertical profile
remains flat if only the distances change. We find that the inferred
vertical profiles are similar to those for the full sample, and the
joint estimate for the slope remains consistent with zero:
$\mathrm{d}\sigma_z(z)/{\mathrm d}z = 0.0 \pm 0.3$ km~s\inv\
kpc\inv. Keeping only stars with $|b| > 60^\circ$ gives similar
results, but leaves only the metal-poor, $\alpha$-enhanced bins with
enough stars do perform the analysis. Thus, we conclude that distance
systematics do not influence the inferred isothermality of the
mono-abundance populations.

\begin{figure}[htbp]
\includegraphics[width=0.5\textwidth,clip=]{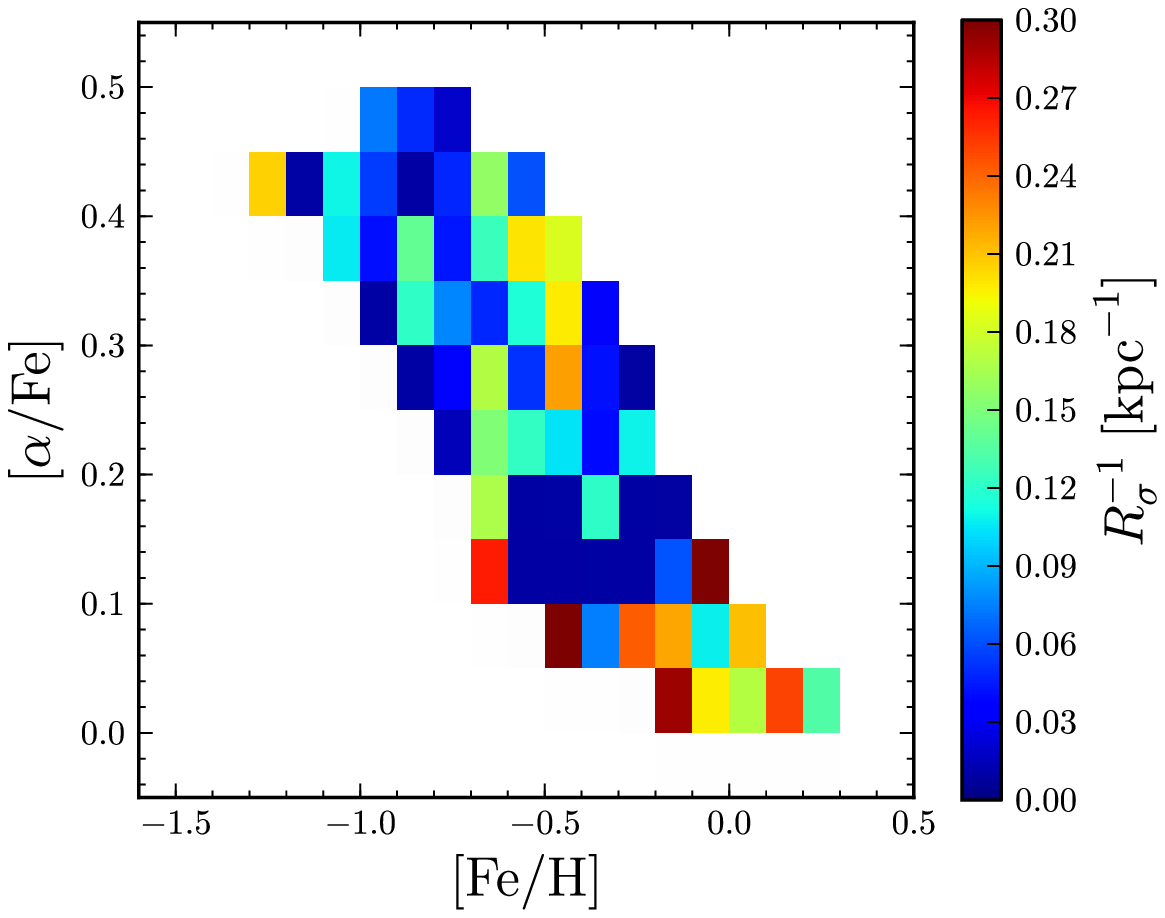}
\includegraphics[width=0.5\textwidth,clip=]{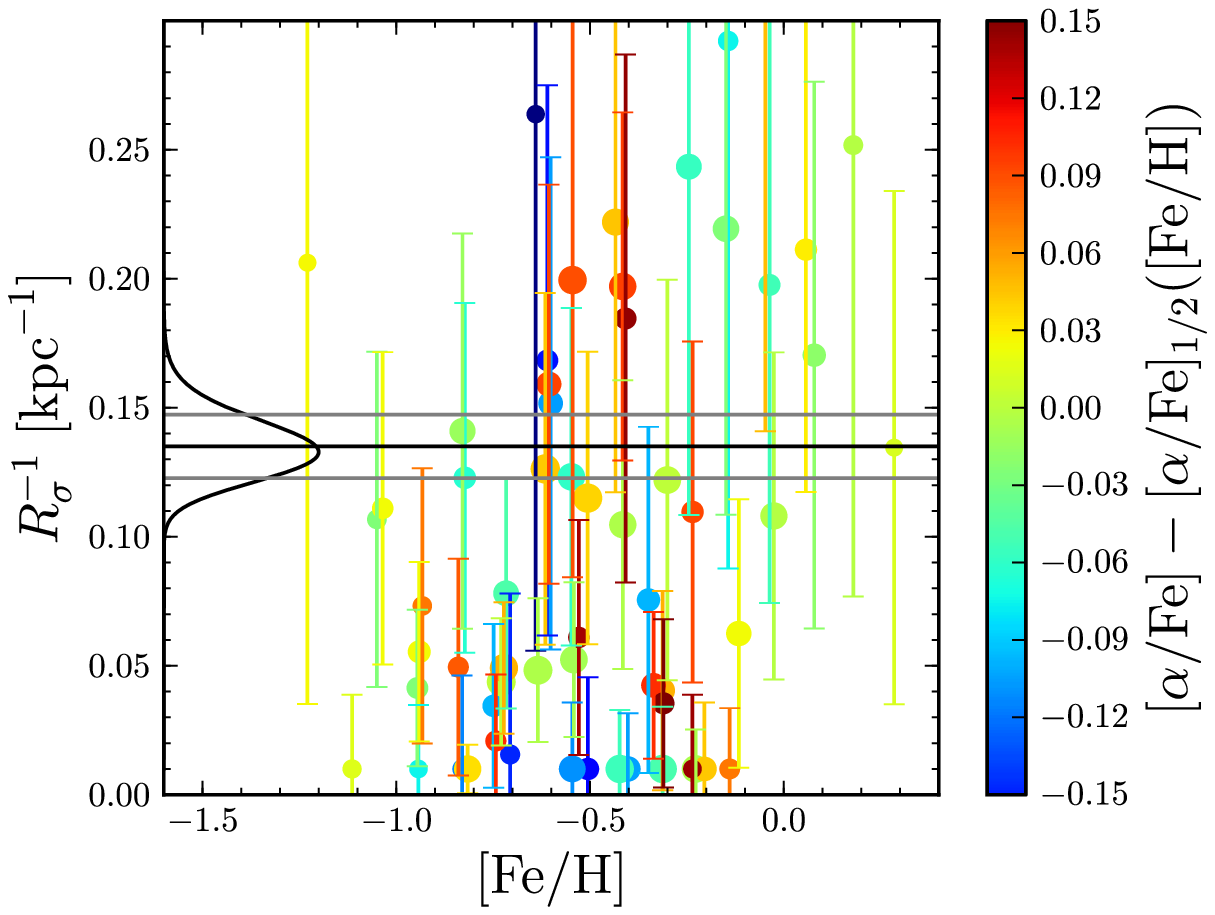}
\caption{Variation of the vertical velocity dispersion with
  Galactocentric radius. Shown is the best-fit inverse radial scale
  length of the vertical velocity dispersion as a function of \feh\
  and \afe\ (left panel) and with error bars as a function of
  metallicity (right panel). Points in the right panel are color-coded
  using \afe\ minus the median \afe\ as a function of \feh. Symbol
  size is proportional to the square root of the number of data points
  used in the point's (\feh,\afe) bin. The distribution shown on the
  vertical axis in the right panel is the joint probability
  distribution for the (mean) inverse radial velocity-dispersion scale
  length.  The black and gray lines show the mean and standard
  deviation, respectively, of this distribution; this mean inverse
  radial dispersion scale length is 0.14$\pm$0.01 kpc$^{-1}$ ($\hs
  \approx 7.1$$\pm$0.5 kpc).}\label{fig:hs}
\end{figure}

\subsection{Radial dependence of the vertical dispersion:
  $\sigma_z(R\, |\, \feh, \afe)$}

\figurename~\ref{fig:hs} presents the dependence of $\sigma_z(z,R\,
|\, \afe , \feh )$ on Galactocentric radius, where the fit parameter
$p_4$ from \eqnname~(\ref{eq:radial}) is represented as an inverse
scale length, $\hs^{-1}$. The left panel shows
$\hs^{-1}(\feh,\afe)$. In contrast to $\sigma_z(\feh,\afe )$, there is
no discernible systematic dependence of $\hs^{-1}$ on \feh\ and
\afe. While $\hs^{-1}$ is not well constrained for any individual
mono-abundance component, a joint analysis combining all of the
individual PDFs for $\hs^{-1}$ yields a PDF for $\hs^{-1}$
characterized by $0.14\pm 0.01$ kpc\inv, corresponding to a typical
outward decrease of $\sigma_z(R)\sim \exp\left (-(R-R_0)/(7.1\pm
  0.5~\mathrm{kpc})\right )$. This outward decrease of the velocity
dispersion, also seen in the luminosity weighted measurements of other
disk galaxies, presumably reflects the outward decrease of the disk
surface-mass density and its corresponding vertical force. We briefly
discuss implications below, but defer dynamical analyses of the
vertical kinematics to a separate paper.

As above, we have also checked the influence of distance systematics
on the inferred $\hs$. Changing the distances by 10\,\% only has a
negligible effect on the inferred $\hs^{-1}$. The joint estimate
remains $\hs^{-1} = 0.14\pm0.01$ kpc\inv\ when increasing the
distances, and changes within the uncertainties to $\hs^{-1} =
0.13\pm0.01$ kpc\inv\ when decreasing the distances. We have also
repeated the analysis using only stars with $|b| > 50^\circ$, finding
$\hs^{-1} = 0.16\pm0.02$ kpc\inv, or $\hs = 6.4\pm0.9$ kpc, which is
consistent with the estimate using the full sample within the
uncertainties.

\subsection{The vertical kinematics as a test of the abundance
  precision}\label{sec:aberrs}

\begin{figure*}
\includegraphics[width=0.5\textwidth,clip=]{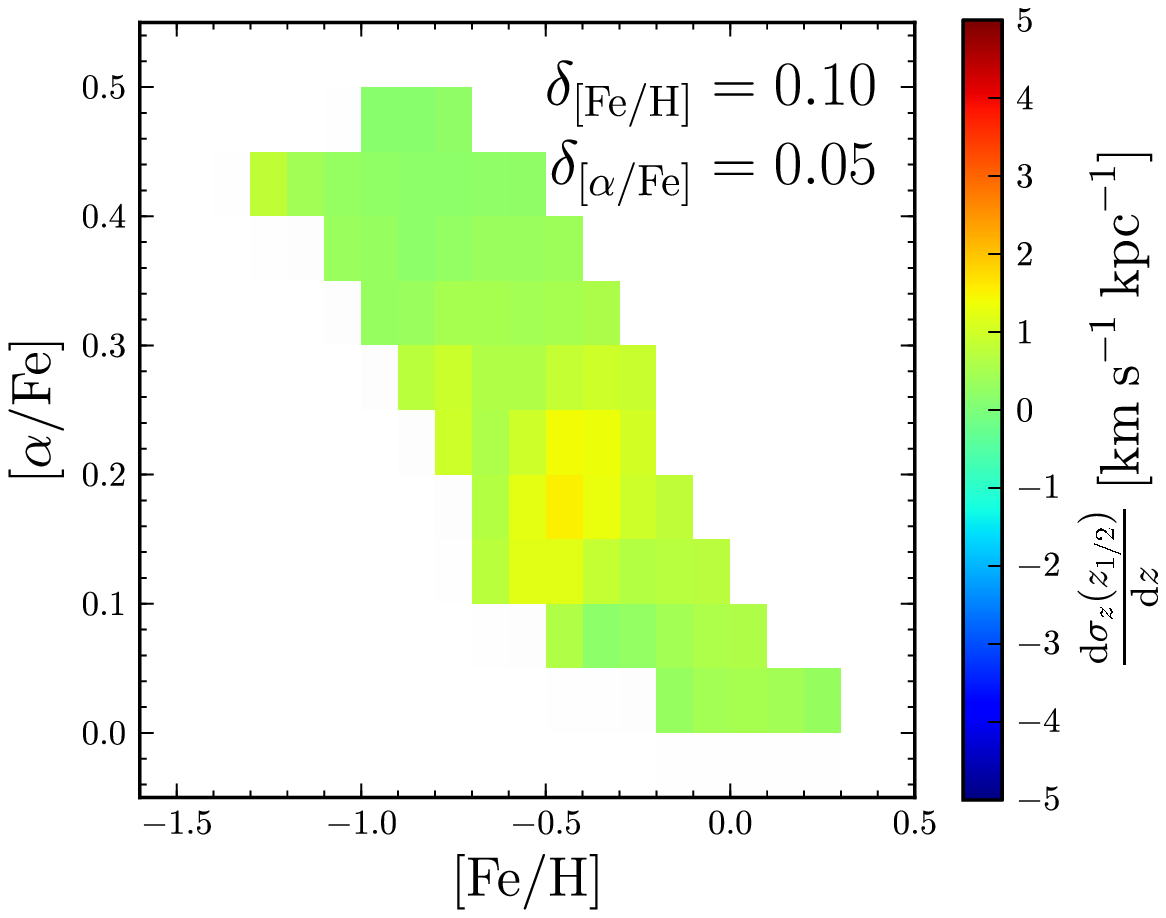}
\includegraphics[width=0.5\textwidth,clip=]{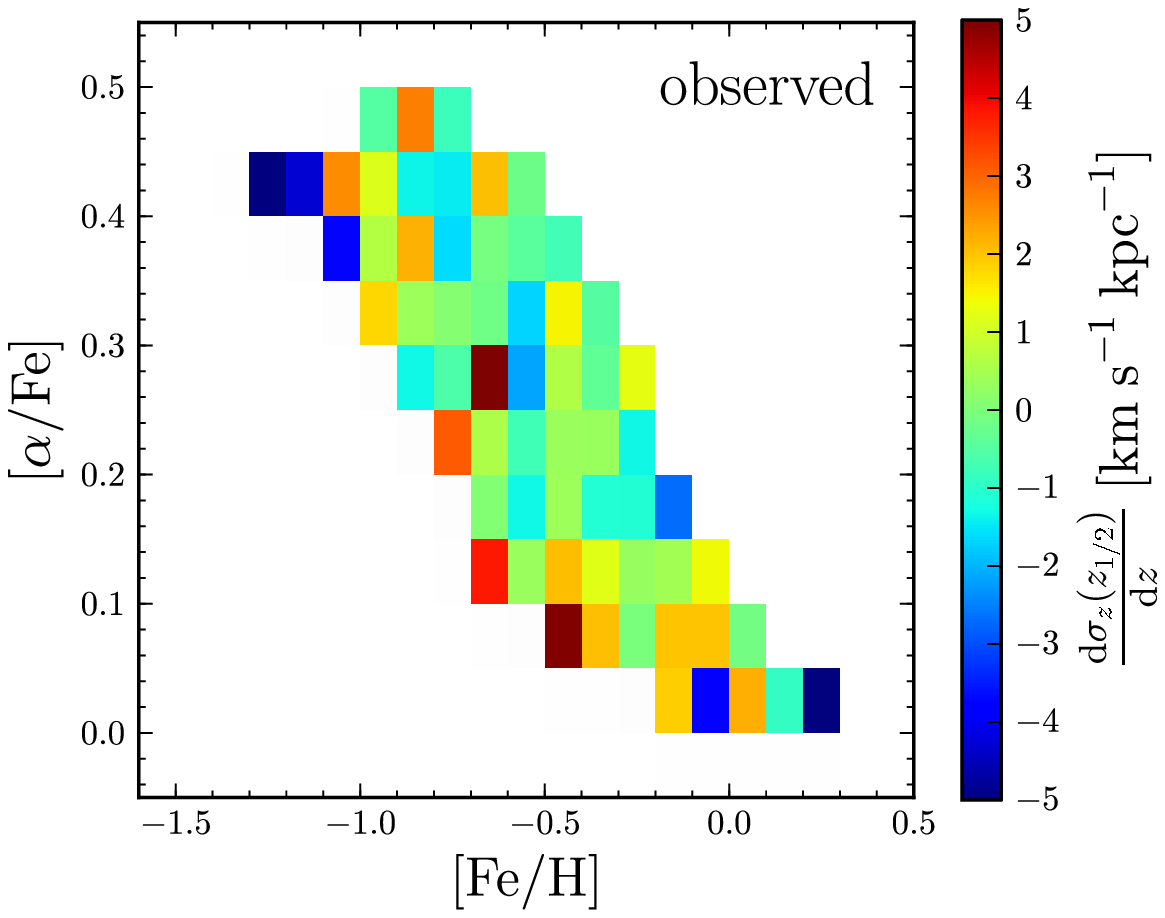}\\
\\
\includegraphics[width=0.5\textwidth,clip=]{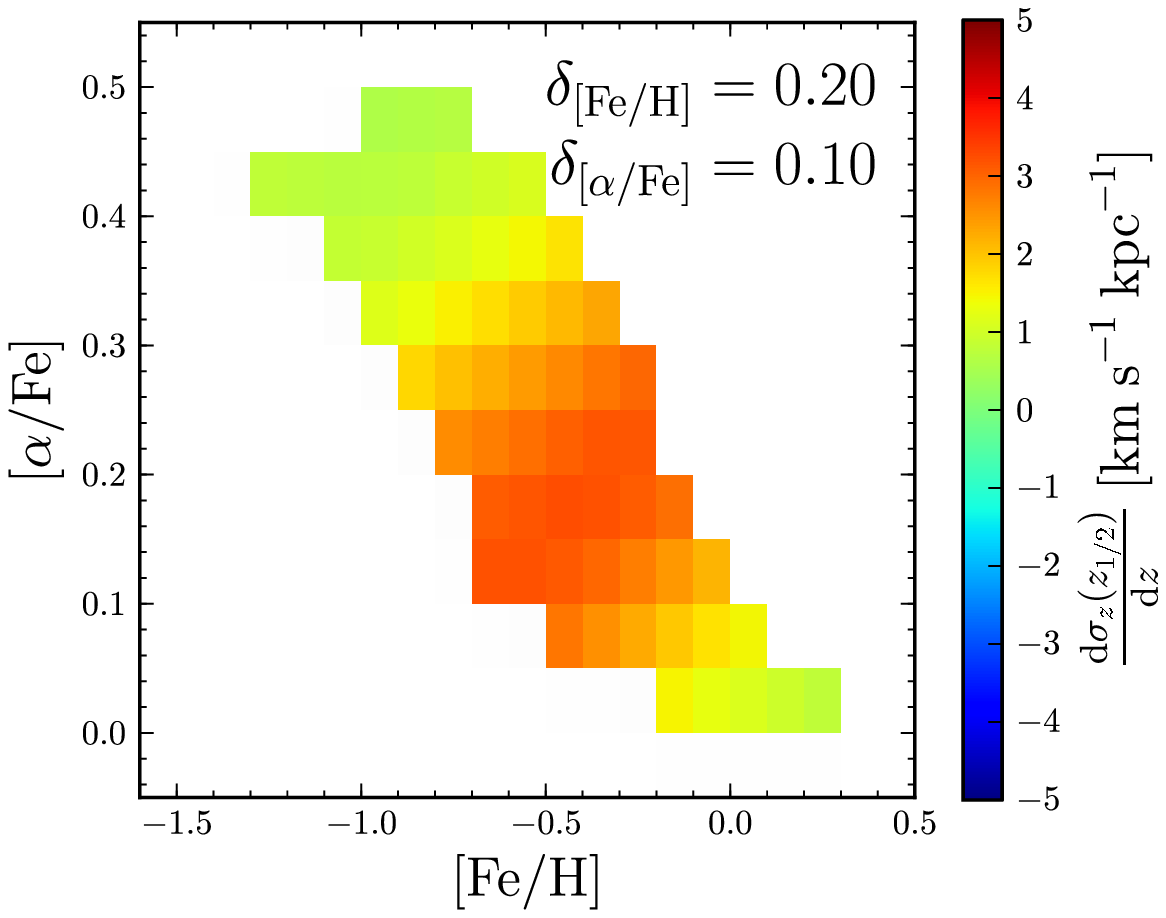}
\includegraphics[width=0.5\textwidth,clip=]{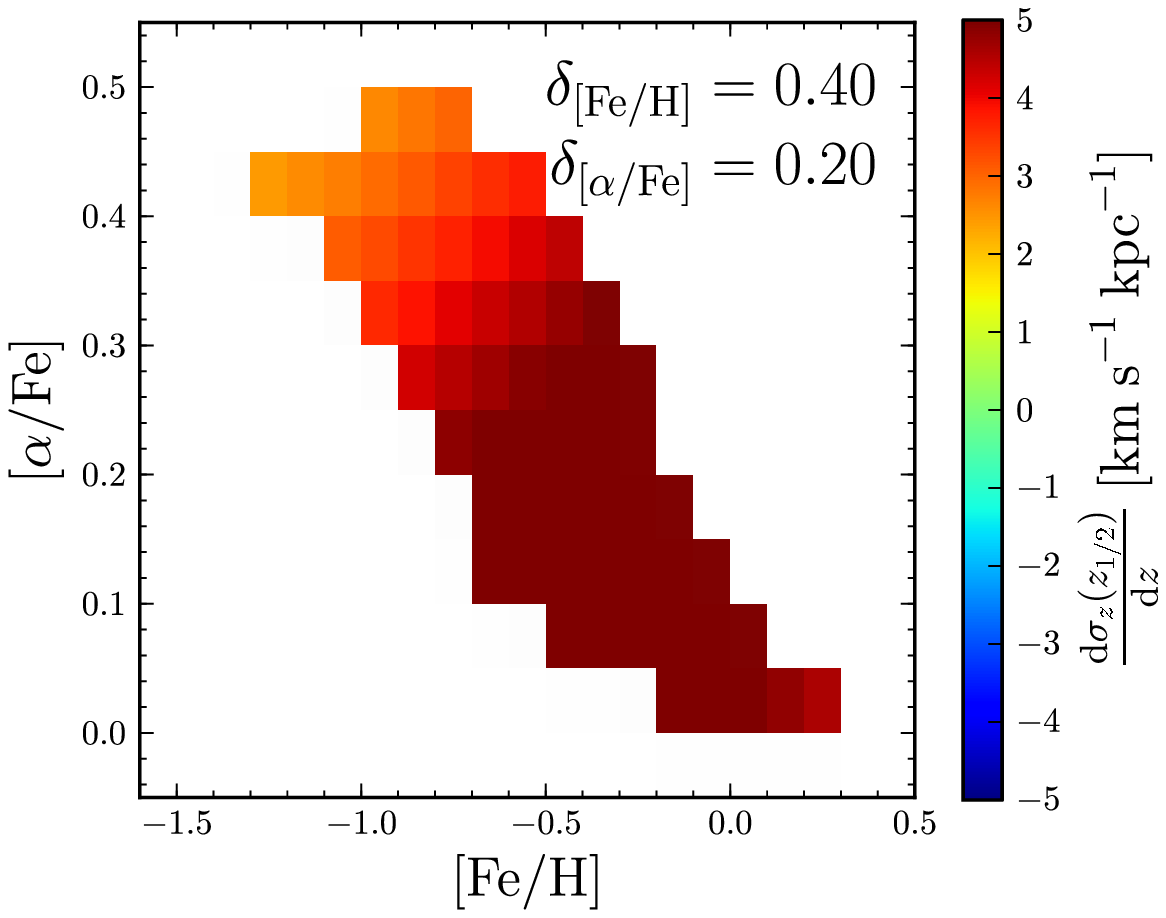}
\caption{Impact of abundance errors on the isothermality of the
  dispersion profile in mono-abundance bins: Three of the panels show
  the \emph{expected} slope of $\sigma_z(z)$ for different magnitudes
  of the \segue\ abundance errors, when assuming that each
  mono-abundance bin is intrinsically isothermal, with velocity
  dispersions given by the top left panel of
  \figurename~\ref{fig:results} and spatial distributions measured by
  \citet{Bovy12a,Bovy12b}. The observed slopes are shown in the top
  right panel. Large abundance errors induce large slopes by `mixing'
  populations of different $\sigma_z$ with proportions that change
  strongly as a function of $|z|$. This is especially the case in the
  transition region between the metal-poor, \afe-enhanced
  sub-populations and sub-populations with solar abundances.  The fact
  that the observed slopes are $\lesssim$ a few km s\inv\ kpc\inv, and
  that there is no discernible trend in the observed slopes in the
  transition region, limits the \segue\ abundance errors to be
  $\delta_{\feh} \approx 0.15$ dex and $\delta_{\afe} \approx 0.07$
  dex (see \figurename~\ref{fig:errs}).}\label{fig:fakeslopes}
\end{figure*}

\begin{figure}[htbp]
\includegraphics[width=0.5\textwidth,clip=]{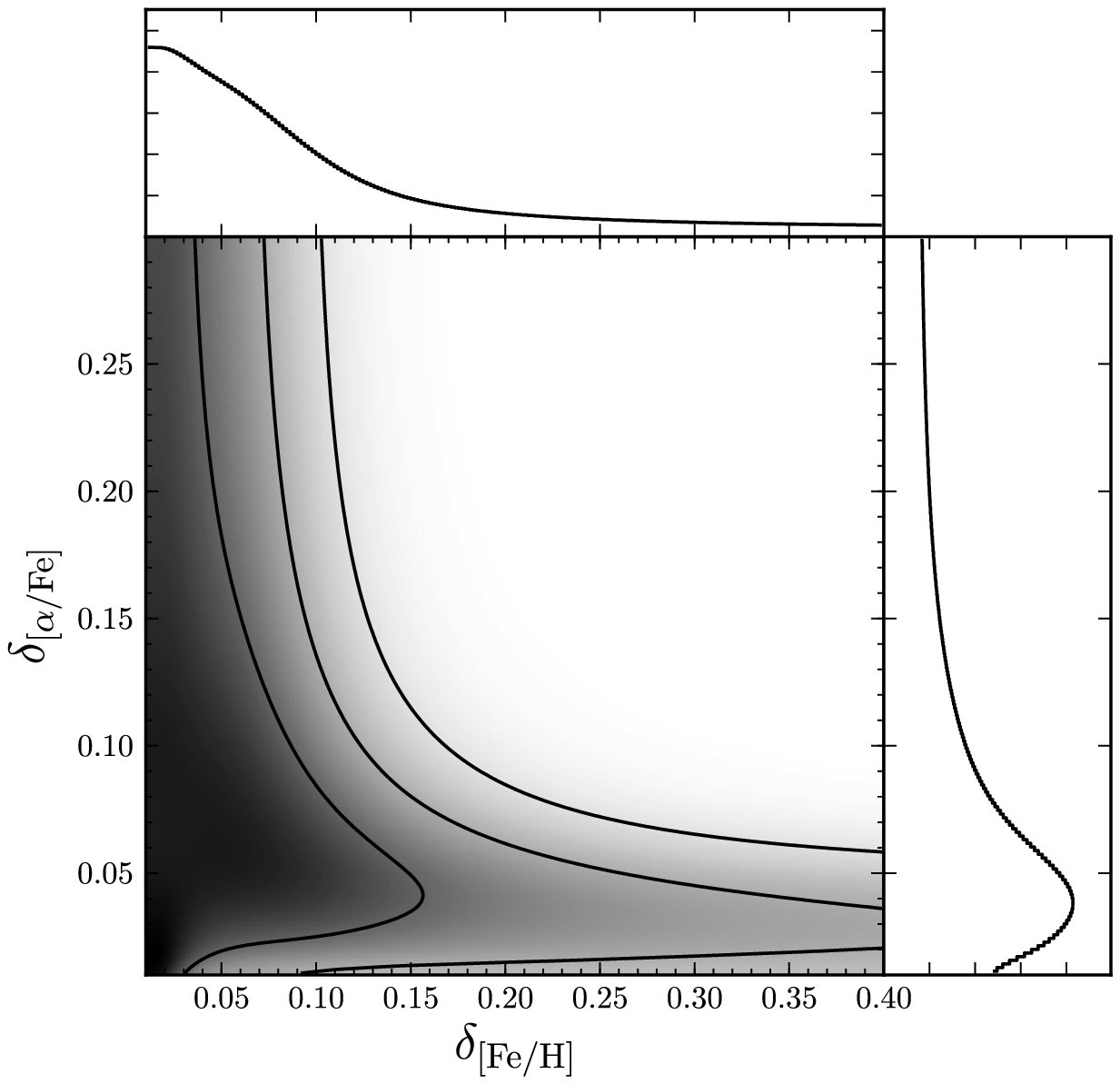}
\includegraphics[width=0.5\textwidth,clip=]{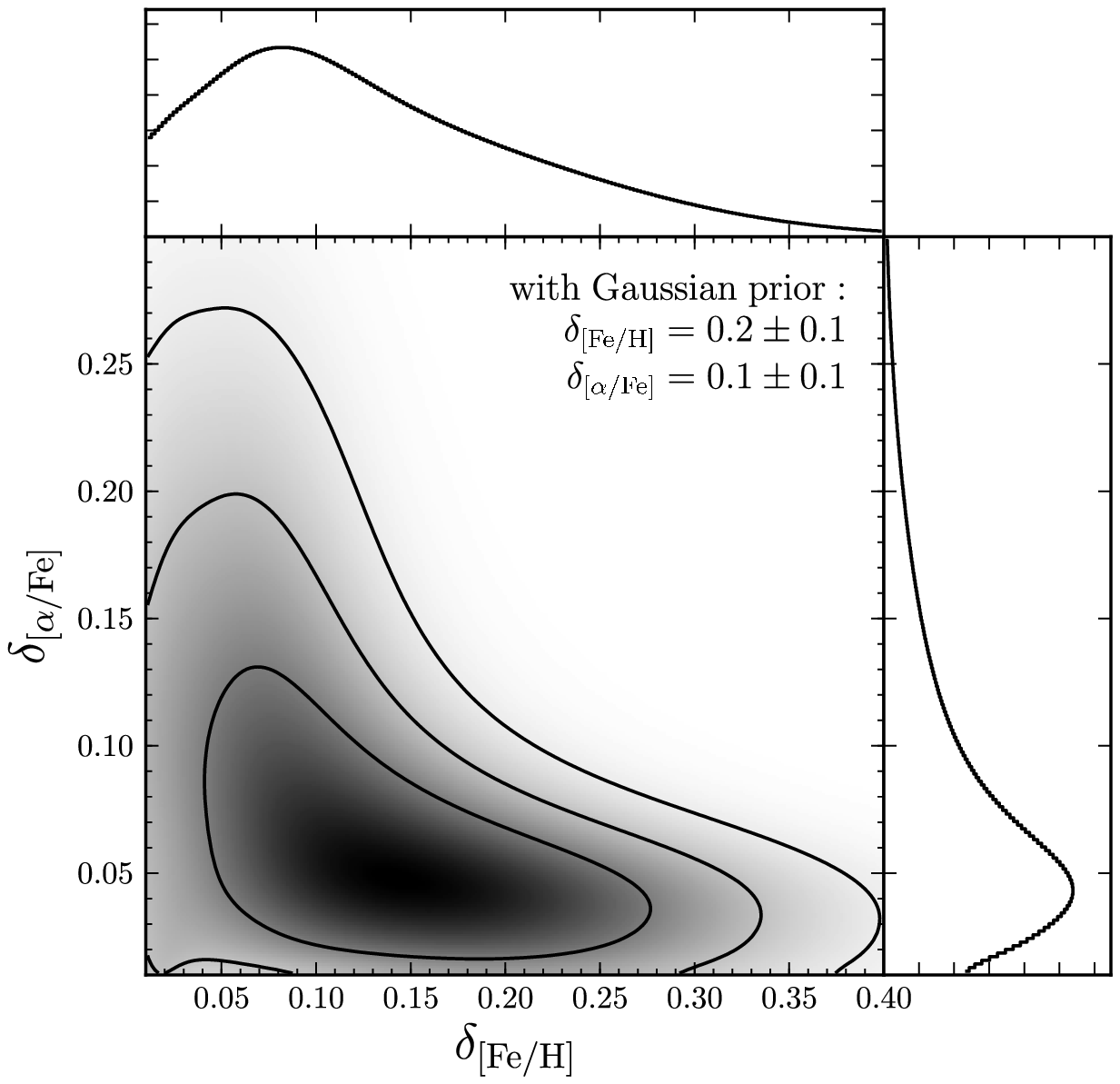}
\caption{Inferred \segue\ abundance uncertainties, assuming that
  mono-abundance sub-populations are intrinsically isothermal. The
  left panel shows the PDF for the abundance uncertainties obtained
  from comparing the observed slopes with the abundance-error induced
  slopes (see \sectionname~\ref{sec:aberrs} and
  \figurename~\ref{fig:fakeslopes}); small uncertainties are
  preferred. The right panel shows the PDF when applying a broad prior
  centered on the \segue-reported abundance uncertainties. The
  preferred abundance uncertainties are still somewhat smaller than
  the \segue-reported errors. We infer that $\delta_{\feh} \approx
  0.15$ dex and $\delta_{\afe} \approx 0.07$ dex.}\label{fig:errs}
\end{figure}

The fact that the data sub-sets we dub `mono-abundance components'
exhibit so nearly an isothermal vertical dispersion profile, $|\langle
{\mathrm d}\sigma_{z}(\zmedian)/{\mathrm d}z(\feh,\afe)\rangle
\lesssim 1$ km s\inv\ kpc\inv, while $\sigma_z(\feh,\afe)$ itself
varies by nearly 40 km s\inv\ among the abundance bins, must set a
constraint on how cleanly our (\feh,\afe) bins separate
abundances, \ie, it allows a completely independent check on the
\emph{precision} (not accuracy) of the \sdss/\segue\ abundance
determinations. As the vertical scale heights of these sub-populations
vary strongly with abundance (\ba), stars with high \afe\ and low
\feh\ will always increase in relative density towards larger $|z|$.
As a consequence, a sub-sample selected nominally by its small \afe\
and high \feh\ values will always have far greater `contamination' at
large $|z|$ from stars with higher \afe\ and lower \feh. As those
contaminants have a substantially higher $\sigma_z$, the dispersion
profile should rise away from the mid-plane if such contamination is
important. This effect is manifested in samples with poor or no
abundance preselection \citep[\eg,][]{Kuijken89a,Fuchs09a}.

To turn the observed isothermality of $\sigma_z(z\, |\, \feh ,\afe)$
into a cross-check on \sdss/\segue\ abundances
\citep{Lee08a,Lee08b,AllendePrieto08a,Lee11a}, we proceed as follows.
We presume that absent any abundance errors, \ie, for the perfect
mono-abundance sub-populations, the dispersion profiles would be
perfectly isothermal. This toy model for $\sigma_z(z)$ is motivated by
the fact that the solution for the vertical Jeans equation in
cylindrical coordinates with the dominant portion of the disk mass at
$|z|<500$~pc (\ie, the limiting case of a mass sheet, with constant
vertical force) and an exponential profile for the tracer density (as
found in \ba), the solution is $\sigma(z)=$~constant. In this context,
any force contribution from a vertically-extended mass distribution
(\eg, the halo), will lead to solutions with $\sigma_z$ increasing with
height, making it unlikely that the error-free dispersion profile
slope is negative (\ie, falling with height). We then assume that the
observed $\sigma_z(\feh ,\afe)$ (top left panel of
\figurename~\ref{fig:results}) and $h_z(\afe ,\feh )$ (Figure 4 in
\ba) are sufficient approximations to the true distribution of these
quantities, and we use the normalizations from \citet{Bovy12b}.
Viewing then the abundance uncertainties $\delta_{\afe}$ and
$\delta_{\feh}$ as free parameters that correlate bins and hence lead
to rising dispersion profiles, we determine for what abundance errors
the predicted and observed slopes of the vertical dispersion profile,
${\mathrm d}\sigma_{z}(\zmedian)/{\mathrm d}z$, or shorter
$\sigma^\prime_z$, match best.  Note that error-induced abundance
correlations will not change the dispersion profile slope in all bins,
as $\sigma_z(\afe ,\feh )$ is a somewhat complex function of
$(\feh,\afe)$.  Specifically, we calculate the PDF for $\delta_{\afe}$
and $\delta_{\feh}$ based on an objective function $\propto \exp\left
  [-0.5 \left
    (\sigma^\prime_{z,obs}-\sigma^\prime_{z,mod}(\delta_{\afe},\delta_{\feh})\right
  )^2 /(\sigma_{\sigma^\prime_{z,obs}})^2 \right]$, where
$\sigma^\prime_{z,mod}(\delta_{\afe},\delta_{\feh})$ is the dispersion
profile slope predicted by the various assumptions on the errors.

\figurename~\ref{fig:fakeslopes} shows the abundance-error-induced
slopes for a few different magnitudes for the abundance uncertainties
and compares these to the observed slopes of
\figurename~\ref{fig:results}, repeated as the top right panel in
\figurename~\ref{fig:fakeslopes}. It is clear that the abundance
uncertainties can certainly not be much bigger than the
\segue-reported uncertainties without inducing large slopes and clear
patterns in the abundance dependence of the slope.

\figurename~\ref{fig:errs} provides the quantitative results of comparing
the abundance-errors-induced slopes with the observed slopes. The left
panel shows the PDF for $\delta_{\afe}$ and $\delta_{\feh}$ using flat
priors on both, while the right panel shows the PDF assuming Gaussian
priors on the error estimates around their nominal value
$\delta_{\afe} = 0.1$ dex \citep{Lee11a} and $\delta_{\feh} = 0.2$ dex
\citep{Smolinski11a}, both with a standard deviation of 0.1 dex.  The
left panel shows that a very clean metallicity separation (\ie, very
small $\feh$ errors) are suggested by the data. If we include the
nominal error priors, we still find that the remarkable
near-isothermality of the mono-abundance dispersion profiles implies
that the abundance precision are somewhat better than the nominal
values, with $\delta_{\feh} \approx 0.15$ dex and $\delta_{\afe}
\approx 0.07$ dex. Note, that this analysis only speaks to the error
\emph{precision}, \ie, the ability to rank stars in their relative
abundance, not to the abundance \emph{accuracy} on any absolute scale.

\section{Discussion}\label{sec:discussion}

\subsection{The simplicity of mono-abundance
  sub-populations}\label{sec:discuss_simple}

The results in this paper show that the vertical motions of
mono-abundance sub-populations of the Milky Way's disk are extremely
simple: they are vertically isothermal to a remarkable degree---we
detect no systematic departure from isothermality at the percent
level---and declining with radius, in a manner that appears consistent
with the decrease in disk surface-mass density (see below). Combined
with the findings in \ba\ that the density of these sub-populations is
well described by a single vertical exponential and a single radial
exponential, this shows that the phase-space structure of
mono-abundance sub-populations is about as simple as it could
be. Chemically-homogeneous sub-components of the disk are individually
simple disk components that come together to create the complex disk
structure that is observed. Typically, this structure has been
decomposed---in light of this paper and of \citet{Bovy12a,Bovy12b}'s
results, perhaps spuriously---into a small number of sub-components
with internal metallicity and velocity gradients.

The fact that the mono-abundance sub-populations are so close to
vertically isothermal is somewhat surprising, given the inherent
complexity in the simple (\feh,\afe) abundance plane induced by the
(presumably) very different star formation histories in different
parts of the disk, and evolutionary processes such as radial migration,
which radially mix stars with very different origins. We expect each
pixel in the (\feh,\afe) abundance plane to contain stars with a
variety of ages and birth locations, with only a rough mean-age
increase with increasing \afe. Whether this a-priori reasonable
picture is inaccurate, because mono-abundance sub-populations are
dominated by a single co-eval population even in the light of
star-formation complexity and radial migration, or whether the Milky
Way relaxes to this simple state, is a question that is best
confronted by more detailed analytic models, simulations or the next
generation of spectroscopic surveys that will determine elemental
abundances beyond \feh\ and \afe, such as \emph{APOGEE}
\citep{Eisenstein12a}.

\begin{figure}[hbtp]
\includegraphics[width=0.5\textwidth]{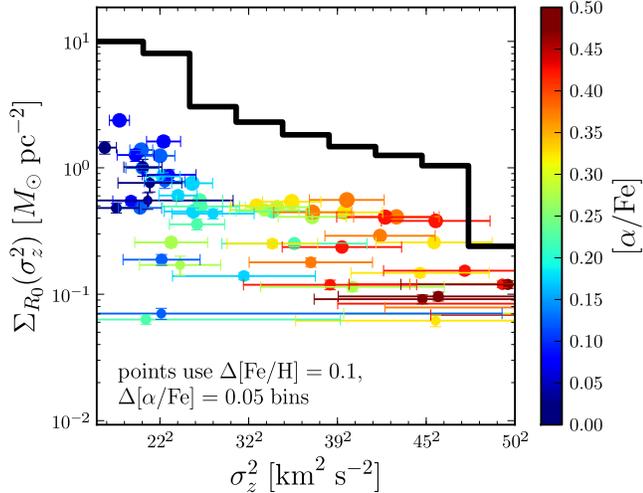}
\caption{Vertical temperature distribution of the stellar disk at
  $\sim\! R_0$: Distribution of stellar surface-mass density at the
  solar radius, $\Sigma_{R_0}(\sigma_z^2)$, as a function of vertical
  velocity dispersion squared, $\sigma_z^2$. The thick black histogram
  shows the total stellar surface-mass contributions in bins in
  $\sigma_z^2$, calculated by summing the total stellar masses of
  sub-populations in \feh\ and \afe. The stellar surface-mass
  densities of individual abundance bins in \feh\ and \afe\ are shown
  as dots, calculated by correcting the observed number counts of
  \segue\ G-type dwarfs into stellar masses using
  stellar-population-synthesis models as described in \citet{Bovy12b},
  with values for $\Sigma_{R_0}(\feh,\afe)$ shown on the $y$-axis.
  The points are color-coded by the value of \afe\ in each bin and the
  size of the points is proportional to the square root of the number
  of data points upon which the kinematic fits are based. The
  temperature distribution exhibits no evidence for a distinct hot
  (``thick'') disk component.}\label{fig:mass_sz_afe}
\end{figure}

The velocity dispersions of the mono-abundance sub-components increase
smoothly from $\sim\!15$ km s\inv\ for the \apoor est populations to
$\gtrsim 50$ km s\inv\ for the \aenhanced est. These extrema are
typical for thin--thick disk decompositions, but the existence of
intermediate populations $\sim\!30$ km s\inv\ is unexpected in the
traditional picture with a thin--thick disk dichotomy. We apply the
procedure of \citet{Bovy12b} for calculating the total stellar
surface-mass density in each mono-abundance bin, correcting the
observed number counts for spectroscopic and
stellar-population-sampling selection effects using
stellar-population-synthesis models (we refer the reader to that paper
for more details on this procedure). This stellar mass is \emph{not}
measured dynamically from the vertical kinematics. This results in the
stellar surface-mass-weighted distribution of the vertical
velocity dispersion or kinetic temperature, shown in
\figurename~\ref{fig:mass_sz_afe}. Together with Figure 2 in
\citet{Bovy12b}, this shows that the distribution of stellar mass in
the Milky Way is monotonically, and seemingly continuously, declining
from cool, thin-disk components to hotter, thicker disk components,
with no gap or bimodality. This, combined with the isothermality of
each mono-abundance component, shows that the intermediate populations
cannot be a mix of the thinner and thicker components, but that they
are real and contribute an amount of stellar mass that is intermediate
between the thinner and thicker components as well. This continuity
rejects the notion of a dichotomy between a thin and a thick disk in
the Milky Way.

We can compare our observational results to some scenarios for
creating the thicker disk components, in particular those where a
pre-existing thin disk is dynamically heated by an infalling
satellite. On this basis, our results reject massive satellite infall
as the dominant disk heating mechanism, as such simulations typically
display a sharp break in the velocity dispersion between the simulated
thin and thick disk \citep{Abadi03a,Brook04a,Villalobos08a}, in
disagreement with our \figurename s~\ref{fig:results} and
\ref{fig:mass_sz_afe}. This, in addition to the density structure in
\ba, points toward an internal mechanism for creating thicker disk
components. Such an internal mechanism could be radial migration
\citep{Loebman11a}. The thick-disk component could also form
internally at early times in a turbulent, clumpy disk
\citep[\eg,][]{Bournaud09a,ForsterSchreiber09a}, although creating the
continuous distribution of vertical temperatures requires that the
disk remains turbulent over a significant fraction of its history
\citep[\eg,][]{Forbes11a}. Whether multiple minor satellite mergers
can plausibly lead to the continuous trends with elemental abundance
of the spatial and kinematic structure in the Milky Way's disk remains
to be seen.

\subsection{The radial dispersion profile and dynamical modeling}

In the simplest model of a razor-thin disk and an exponential tracer
density, solution of the vertical Jeans equation yields the disk
surface-mass density, $\Sigma(R)$, in terms of the vertical velocity
dispersion $\sigma_z$ and the vertical scale height $h_z$ of the
tracer population: $\Sigma(R) \propto \sigma_z^2 / h_z$. Assuming that
the tracer's density does not flare, this implies that the mass scale
length of the disk is equal to half of the exponential scale length of
the velocity dispersion, and that the dispersion scale length has no
abundance dependence. We indeed do not observe any abundance
dependence in the inferred $\hs\inv$ in \figurename~\ref{fig:hs}, and
\ba\ found no evidence for a flare in the tracers' density
profile. Thus, we obtain an estimate of the disk's mass scale length of
$\approx 3.5$ kpc from our combined estimated of the dispersion scale
length of $\hs = 7.1 \pm 0.5$ kpc. The addition of the dark matter
halo, which locally has an equivalent exponential scale length
$\gtrsim 5$ kpc, means that this is likely an overestimate, such that
we expect the scale length to be $\lesssim 3.5$ kpc.

Rather than trying to estimate the halo correction to the disk scale
length as estimated above, we defer a proper dynamical analysis to a
separate paper. The simplicity of the mono-abundance sub-populations
makes them excellent tracers of the local vertical potential, because
their spatial and kinematic simplicity indicates that their
phase-space distribution function is simple. This removes the common
difficulty in modeling and marginalizing over the phase-space
distribution of a kinematic tracer sample when doing dynamical
inference \citep[\eg,][]{Bovy10a}. The large number of independent
mono-abundance sub-populations that all trace the same underlying
gravitational potential will permit for extensive cross-checks. The
abundance independence of the inferred dispersion scale length in this
paper is a first example of this, and indicates how fruitful such an
approach may be.

\section{Summary}\label{sec:conclusion}

This paper obtains the following results:

$\bullet$ We modeled the vertical velocity dispersion, $\sigma_z(z,R)$,
and its spatial dependence of mono-abundance (\feh,\afe) slices of the
\segue\ G-dwarf sample. We found that all mono-abundance components
exhibit nearly isothermal kinematics in the vertical direction, and a
slow outward decrease in the radial direction:
\[ \sigma_z\bigl (z,R\,|\, \afe,\feh \bigr )\approx \sigma_z\bigl
(\afe,\feh \bigr )\]\[\qquad \qquad \qquad \qquad \qquad \qquad \times\exp\bigl (-(R-R_0)/7\, \mathrm{ kpc}\bigr
)\,.\] Each mono-abundance component is isothermal to within about 3
km s$^{-1}$ kpc$^{-1}$. The mean vertical $\sigma_z$ gradient is only
$0.3\pm0.2$ km s\inv\ kpc\inv, such that the mono-abundance
populations are consistent with isothermality at the level of 1\,\%,
when compared to the range of velocity dispersions present in this
sample.

$\bullet$ We find a smooth variation in characteristic velocity
dispersion from $\sim\!15$ km s\inv\ for chemically-young stars with
solar abundances to $\gtrsim 50$ km s\inv\ for metal-poor,
\afe-enhanced stars that are presumably old. The mass-weighted
distribution of vertical kinetic temperatures ($\propto \sigma_z^2$)
is a continuous and monotonically-declining function, similar to the
mass-weighted distribution of scale heights found in
\citet{Bovy12b}. This, and in particular the existence of isothermal,
intermediate-dispersion ($\sim\!30$ to 40 km s\inv) sub-populations,
shows that the mono-abundance sub-populations cannot be the mix of a
thin, cool disk component and a thick, hotter component. The data
therefore reject the notion of a thin--thick-disk dichotomy.

$\bullet$ The remarkable degree of isothermality of the mono-abundance
components also allows us to independently verify the
\segue\ abundance precision, as it limits the range of abundances
actually present within a bin: we find that $\delta_{\afe}\approx
0.07$ dex and $\delta_{\feh}\approx 0.15$ dex.

$\bullet$ The radial dependence of the vertical velocity dispersion
can be described by an exponential. The exponential scale length of
this radial dispersion profile has no obvious abundance dependence, as
expected when a massive disk contributes significantly to the
gravitational potential. The exponential decline of $\sigma_z^2$
presumably traces the radial decline of the disk's surface-mass
density, but the quality of the data demands careful dynamical
modeling beyond the scope of this paper. The simplicity of the
vertical phase-space structure of the mono-abundance sub-populations
will allow detailed dynamical modeling of the Milky Way's potential
near the disk plane.

\acknowledgements It is a pleasure to thank the anonymous referee, Dan
Foreman-Mackey, \v{Z}eljko Ivezi\'{c}, Mario Juri\'{c}, Chao Liu,
Scott Tremaine, and Glenn van de Ven for helpful comments and
assistance. We thank the \segue\ team for their efforts in producing
the \segue\ data set.  Support for Program number HST-HF-51285.01-A
was provided by NASA through a Hubble Fellowship grant from the Space
Telescope Science Institute, which is operated by the Association of
Universities for Research in Astronomy, Incorporated, under NASA
contract NAS5-26555.  J.B. and D.W.H. were partially supported by NASA
(grant NNX08AJ48G) and the NSF (grant AST-0908357). D.W.H. is a
research fellow of the Alexander von Humboldt Foundation of
Germany. J.B. and H.W.R. acknowledge partial support from SFB 881
funded by the German Research Foundation DFG. T.C.B. and Y.S.L
acknowledge partial support by grants PHY 02-16783 and PHY 08-22648:
Physics Frontiers Center/Joint Institute for Nuclear Astrophysics
(JINA), awarded by the U.S. National Science Foundation.

Funding for the SDSS and SDSS-II has been provided
by the Alfred P. Sloan Foundation, the Participating Institutions, the
National Science Foundation, the U.S. Department of Energy, the
National Aeronautics and Space Administration, the Japanese
Monbukagakusho, the Max Planck Society, and the Higher Education
Funding Council for England. The SDSS Web Site is
http://www.sdss.org/.


\begin{thebibliography}{}

\bibitem[Abadi \etal(2003)]{Abadi03a}
  Abadi,~M.~G., Navarro,~J.~F., Steinmetz,~M., \& Eke,~V.~R., 2003, \apj, 591, 499
\bibitem[Abazajian \etal(2009)]{Abazajian09a}
  Abazajian,~K.~N., Adelman-McCarthy,~J.~K., Ag\~{u}eros,~M.~A., \etal\ 2009, \apjs, 182, 543
\bibitem[Allende Prieto \etal(2008)]{AllendePrieto08a}
  Allende Prieto,~C., Sivarani,~T., Beers,~T.~C., \etal\ 2008, \aj, 136, 2070
\bibitem[Bensby \etal(2003)]{Bensby03a}
  Bensby,~T., Feltzing,~S., \& Lundstr\"{o}m,~I. 2003, \aap, 410, 527
\bibitem[Bond \etal(2010)]{Bond10a}
  Bond,~N., Ivezi\'{c},~\v{Z}., Sesar,~B., \etal\ 2010, \apj, 716, 1
\bibitem[Bournaud \etal(2009)]{Bournaud09a}
  Bournaud,~F., Elmegreen,~B.~G., \& Martig,~M. 2009, \apjl, 707, L1
\bibitem[Bovy \etal(2009)]{Bovy09b}
  Bovy,~J., Hogg,~D.~W., \& Rix, H.-W. 2009, \apj, 704, 1704
\bibitem[Bovy \etal(2010)]{Bovy10a}
  Bovy,~J., Murray,~I., \& Hogg,~D.~W.\ 2010, \apj, 711, 1157
\bibitem[Bovy \etal(2012b)]{Bovy12b}
  Bovy,~J.,~Rix,~H.-W., Hogg,~D.~W., \etal\ 2012b, \apj, 751, 131
\bibitem[Bovy \etal(2012a)]{Bovy12a}
  Bovy,~J.,~Rix,~H.-W., Liu,~C., \etal\ 2012a, \apj, 753, 148
\bibitem[Brook \etal(2004)]{Brook04a}
  Brook,~C.~B., Kawata,~D., Gibson,~B.~K., \& Freeman,~K.~C., \apj, 612, 894
\bibitem[Chen \etal(2001)]{Chen01a}
  Chen,~B., Stoughton,~C., Smith,~J.~A., \etal\ 2001, \apj, 553, 184
\bibitem[Chiba \& Beers(2000)]{Chiba00a}
  Chiba,~M.~\& Beers,~T.~C. 2000, \aj, 119, 2843
\bibitem[Eisenstein \etal(2012)]{Eisenstein12a}
  Eisenstein,~D.~J., Weinberg,~D.~H., Agol,~E., \etal\ 2012, \aj, 142, 72
\bibitem[Feltzing \etal(2003)]{Feltzing03a}
  Feltzing,~S., Bensby,~T., \& Lundstr\"{o}m,~I. 2003, \aap, 397, 1
\bibitem[Flynn \& Fuchs(1994)]{Flynn94a}
  Flynn,~C.~\& Fuchs,~B. 1994, \mnras, 270, 471
\bibitem[Forbes \etal(2011)]{Forbes11a}
  Forbes,~J., Krumholz,~M., \& Burkert,~A. 2012, \apj, 754, 48
\bibitem[Foreman-Mackey \etal(2012)]{FM12a} Foreman-Mackey,~D.,
  Hogg,~D.~W., Lang,~D., \& Goodman, J.\ 2012, arXiv:1202.3665v2 [astro-ph.IM]
\bibitem[F{\"o}rster Schreiber \etal(2009)]{ForsterSchreiber09a}
  F{\"o}rster Schreiber,~N.~M., Genzel,~R., Bouch{\'e},~N., \etal\ 2009, \apj, 706, 1364
\bibitem[Fuchs \etal(2009)]{Fuchs09a}
  Fuchs,~B., Dettbarn,~C., Rix,~H.-W., \etal\ 2009, \aj, 137, 4149
\bibitem[Fuhrmann(1998)]{Fuhrmann98a}
  Fuhrmann,~K. 1998, \aap, 338, 161
\bibitem[Goodman \& Weare(2010)]{Goodman10a}
  Goodman,~J.~\& Weare,~J., 2010, Comm.\ App.\ Math.\ and Comp.\ Sci., 5, 65
\bibitem[Ivezi\'{c} \etal(2008)]{Ivezic08a}
  Ivezi\'{c},~\v{Z}., Sesar,~B., Juric,~M., \etal\ 2008, \apj, 684, 287
\bibitem[Juri\'{c} \etal(2008)]{Juric08a}
  Juri\'{c},~M., Ivezi\'{c},~\v{Z}., Brooks,~A., \etal\ 2008, \apj, 673, 864
\bibitem[Kuijken \& Gilmore(1989)]{Kuijken89a}
  Kuijken,~K.~\& Gilmore,~G. 1989, \mnras, 239, 605
\bibitem[Lee \etal(2011a)]{Lee11a}
  Lee,~Y.~S., Beers,~T.~C., Allende Prieto,~C., \etal\ 2011a, \aj, 141, 90
\bibitem[Lee \etal(2011b)]{Lee11b}
  Lee,~Y.~S., Beers,~T.~C, An,~D., \etal\ 2011b, \apj, 738, 187
\bibitem[Lee \etal(2008a)]{Lee08a}
  Lee,~Y.~S., Beers,~T.~C., Sivarani,~T., \etal\ 2008a, \aj, 136, 2022
\bibitem[Lee \etal(2008b)]{Lee08b}
  Lee,~Y.~S., Beers,~T.~C., Sivarani,~T., \etal\ 2008b, \aj, 136, 2050
\bibitem[Liu \& van de Ven(2012)]{Liu12a}
  Liu,~C.~\& van de Ven,~G.~2012, \mnras, in press, arXiv:1201.1635 
\bibitem[Loebman \etal(2011)]{Loebman11a}
  Loebman,~S.~R., Ro{\v s}kar,~R., Debattista,~V.~P., \etal\ 2011, \apj, 737, 8
\bibitem[Munn \etal(2004)]{Munn04a}
  Munn,~J.~A., Monet,~D.~G., Levine,~S.~E., \etal\ 2004, \aj, 127, 3034
\bibitem[Nordstr\"{o}m \etal(2004)]{Nordstroem04a}
  Nordstr\"{o}m,~B., Mayor,~M., Andersen,~J., \etal\ 2004, \aap, 418, 989
\bibitem[Prochaska \etal(2000)]{Prochaska00a}
  Prochaska,~J.~X., Naumov,~S.~O., Carney,~B.~W., McWilliam,~A., \& Wolfe,~A.~M. 2000, \aj, 120, 2513
\bibitem[Schlesinger \etal(2010)]{Schlesinger10a}
  Schlesinger,~K.~J., Johnson,~J.~A., Rockosi,~C.~M., \etal\ 2010, \apj, 719, 996
\bibitem[Sch\"{o}nrich \& Binney(2009)]{Schoenrich08a}
  Sch\"{o}nrich,~R.~\& Binney,~J.~J. 2009, \mnras, 396, 203
\bibitem[Sch\"{o}nrich \etal(2010)]{Schoenrich10a}
  Sch\"{o}nrich,~R., Binney,~J.~J., \& Dehnen,~W.~2010, \mnras, 403, 1829
\bibitem[Smolinski \etal(2011)]{Smolinski11a}
  Smolinski,~J.~P., Lee,~Y.~S., Beers,~T.~C., \etal\ 2011, \aj, 141, 89
\bibitem[Villalobos \& Helmi(2008)]{Villalobos08a}
  Villalobos,~A.~\& Helmi,~A., 2008, \mnras, 391, 1806
\bibitem[Yanny \etal(2009)]{Yanny09a}
  Yanny,~B., Rockosi,~C., Newberg,~H.~J., \etal\ 2009, \aj, 137, 4377
\end{thebibliography}
\end{document}